\documentclass[a4paper,10pt]{article}

\usepackage[utf8]{inputenc}
\usepackage{amssymb}
\usepackage{amsmath}
\usepackage{graphicx}
\usepackage{xcolor}
\usepackage{url}
\usepackage[round,sort]{natbib}

\usepackage{algorithm}
\usepackage{algpseudocode}
\usepackage{algorithmicx}
\usepackage[left=2cm,top=2.5cm,right=2cm,bottom=2.5cm]{geometry}

\graphicspath{{Figures/}}

\title{libEMM: A fictious wave domain 3D CSEM modelling library bridging sequential and parallel GPU implementation}

\author{Pengliang Yang$^1$\\
  $^1$ School of Mathematics, Harbin Institute of Technology, 150001, Harbin, China\\
  E-mail: ypl.2100@gmail.com
}

\begin{document}

\maketitle

\begin{abstract}
  This paper delivers a software - \verb|libEMM| - for 3D controlled-source electromagnetics (CSEM) modelling in fictitious wave domain, based on the newly developed high-order finite-difference time-domain (FDTD) method on non-uniform grid. The numerical simulation can be carried out over a number of parallel processors using MPI-based high performance computing architecture. The FDTD kernel coded in C has been parallelized with OpenMP for speedup using local shared memory. In addition, the software features a GPU implementation of the same algorithm based on CUDA programming language, which can be cross-validated and compared in terms of efficiency. A perspective of \verb|libEMM| on the horizon is its application to 3D CSEM inversion in land and marine environment.
\end{abstract}

\section{Introduction}

Controlled-source electromagnetics (CSEM) has been a well established technology in detecting hydrocarbon bearing formations to do geophysical explorations \citep{chave1982on,constable1986offshore}. CSEM has been applied to air-bone \citep{yin2015review} and cross-well \citep{alumbaugh1997three} geometries, in land  \citep{ward1988electromagnetic,zhdanov1994geoelectrical,grayver2014geo} and/or marine environment \citep{eidesmo2002sbl,ellingsrud2002remote,constable2010,macgregor2014mcsem}. These applications are based on the high sensitivity of the low-frequency electromagnetic signals to detect high resistivity contrast in the subsurface. It makes CSEM an ideal tool for prospect de-risking \citep{macgregor2007derisking} when a decision on well-placement has to be made prior to drilling \citep{meju2019simple}.

CSEM inversion of electromagnetic data is a powerful imaging approach to distinguishing the strong resistive anomalies, thus helping to decipher the potential hydrocarbon distribution in the Earth. CSEM inversion relies on repeated application of  3D numerical modelling engine based on diffusive Maxwell equation. To do that, different methods have been developed, i.e., frequency-domain finite-difference method \citep{newman1995frequency,smith1996conservative1,mulder2006multigrid,streich20093d}, frequency-domain finite-element method \citep{li20072d,da2012finite,key2016mare2dem,rochlitz2019custem}, time-domain finite-difference method \citep{oristaglio1984diffusion,wang1993finite,Taflove_2005_CEF}.

A number of open source codes have been released for CSEM modelling in public domain, such as \verb|SimPEG| \citep{cockett2015simpeg}, \verb|MARE2DEM| \citep{key2016mare2dem},  \verb|PETGEM| \citep{castillo2018petgem}, \verb|custEM| \citep{rochlitz2019custem} and \verb|emg3d| \citep{werthmuller2019emg3d}. Among them, \verb|MARE2DEM| in Fortran is dedicated to 2D modelling and inversion, while the remaining four  are built in python working in 3D.  Both \verb|SimPEG| and \verb|emg3d| allow forward simulation and estimation of gradient in inversion mode.
The forward simulation of \verb|SimPEG| \citep{cockett2015simpeg} and \verb|emg3d| \citep{werthmuller2019emg3d} relies on finite-volume meshes, while \verb|PETGEM| \citep{castillo2018petgem} and \verb|custEM| \citep{rochlitz2019custem} are designed for accurate modelling in complicated models with finite elements.
All these software packages are based on a frequency domain approach by solving the linear system after discretization of the frequency domain Maxwell equation. Some of them support the output of EM field as time series, by first modelling at judiciously selected frequencies and then Fourier transforming the frequency domain field into time domain \citep{rochlitz2021evaluation}.

The main contribution of this paper is to deliver a time-domain 3D CSEM modelling code - \verb|libEMM|. The code implements our newly developed high-order FDTD scheme running on both uniform and nonuniform (NU) staggered grid \citep{Yang_2023_HFDNU}, using the efficient fictitious wave domain approach to do 3D CSEM modelling \citep{Mittet_2010_HFD}. The key difference between \verb|libEMM| and other softwares is the computation of the EM field by a direct discretization of the converted Maxwell equation in fictitious time domain. Significant performance improvement in  accuracy and computational efficiency of this method have been achieved compared to  the commonly used 2nd order scheme over nonuniform grid \citep{Yang_2023_HFDNU}.

This paper focuses on the computer implementation in a software framework rather than numerical justification of the method in terms of accuracy. Besides the implementation using message passing interface (MPI) over a number of distributed CPU cores, the paper also presents a number of key techniques to achieve multithread GPU parallelization. This is also the first time that GPU implementation of this method is extended from uniform grid \citep{Yang_2021_GPU_CSEM} to nonuniform grid, and released in an open source software, in the hope of wide adoption. Consequently, the coding of \verb|libEMM| involves programming using C and Nvidia compute unified device architecture (CUDA), parallelized with MPI. The modelling job can then be launched using shell script on high performance computing cluster.

The paper is organized in the following. First we briefly review the methodology of CSEM modelling in fictitious wave domain. Then we present key ingredients to make an effective implementation, including the boundary conditions (the top air-water boundary condition and the absorbing boundary condition on other sides), the time stepping, the medium homogenization and the optimal grid generation.
To facilitate CSEM modelling on GPU-enabled hardware, we also list several strategies  to optimize code using CUDA programming language. After that, we catch the main features of \verb|libEMM| for an easy-take in terms of software usage. Then, application examples are given with performance analysis. A final remark will be made towards the potential applications in real 3D CSEM inversion.

\section{CSEM modelling in fictitious wave domain}

\subsection{The diffusive Maxwell equation}

The CSEM physics is governed by the diffusive Maxwell equation (consisting of Faraday's law and Amp\`{e}re's law), which reads in the frequency domain
\begin{equation}
 \begin{cases}
 \nabla \times \mathbf{E} -\mathrm{i}\omega\mu  \mathbf{H} = -\mathbf{M} \\
 -\nabla \times \mathbf{H} +\sigma \mathbf{E}=-\mathbf{J}
 \end{cases},
\end{equation}
where $\mathbf{E}=(E_x,E_y,E_z)^T$ and $\mathbf{H}=(H_x, H_y, H_z)^T$ are the electrical and magnetic fields; $\mathbf{J}=(J_x, J_y, J_z)^T$ and $\mathbf{M}=(M_x, M_y, M_z)^T$ are the electrical and magnetic sources. Here, we slightly abuse the same notation to denote the same quantity when switching between time domain and frequency domain, where the convention of Fourier transform $\partial_t \leftrightarrow -\mathrm{i}\omega$ has been adopted.
The magnetic permeability is $\mu$. The conductivity is a symmetric $3\times 3$ 
tensor:
\begin{equation}
  \sigma = \left[\begin{array}{lll}
    \sigma_{xx} & \sigma_{xy} & \sigma_{xz}\\
    \sigma_{yx} & \sigma_{yy} & \sigma_{yz}\\
    \sigma_{zx} & \sigma_{zy} & \sigma_{zz}
    \end{array}\right]
\end{equation}
in which $ \sigma_{ij}=\sigma_{ji}$, $i,j\in\{x,y,z\}$. The isotropic medium means only the diagonal elements of the conductivity
tensor are non-zeros and the same in all directions:
$\sigma_{xx}=\sigma_{yy}=\sigma_{zz}$; $\sigma_{ij}=0, i\neq j $.
The vertical transverse isotropic (VTI) medium still has only the diagonal 
elements, that is, $ \sigma_h:=\sigma_{xx}=\sigma_{yy},\quad \sigma_v=\sigma_{zz}$, where $\sigma_h$ and $\sigma_v$ stand for horizontal conductivity and vertical conductivity, respectively. The resistivity is also frequently used in CSEM system which is the inverse of the conductivity ($\rho_{ij}=1/\sigma_{ij}$).

\subsection{Fictitious wave domain modelling}

By defining a fictitious di-electrical permittivity $\epsilon$ such that $\sigma = 2\omega_0 \varepsilon$, and multiplying the second equation with $\sqrt{-\mathrm{i}\omega/2\omega_0}$, \citep{Mittet_2010_HFD}  transformed the equation into
\begin{equation}\label{eq:mittet}
\begin{cases}
\nabla\times \mathbf{E}'-\mathrm{i}\omega'\mu \mathbf{H}' = -\mathbf{M}'\\
-\nabla\times\mathbf{H}'-\mathrm{i}\omega'\varepsilon \mathbf{E}' = -\mathbf{J}'
\end{cases} 
\end{equation}
where we introduce a prime to define the resulting wave domain fields via
\begin{equation}\label{eq:correspondence}
    \mathbf{E}'=\mathbf{E},\;
    \mathbf{H}'= \sqrt{\frac{-\mathrm{i}\omega}{2\omega_0}} \mathbf{H},\;
    \mathbf{M}'= \mathbf{M},\;
    \mathbf{J}'= \sqrt{\frac{-\mathrm{i}\omega}{2\omega_0}} \mathbf{J},\;
    \omega' = (1+\mathrm{i})\sqrt{\omega\omega_0}.
\end{equation}
Equation \eqref{eq:mittet} may be transformed into time domain as
\begin{equation}
\begin{cases}
\nabla\times \mathbf{E}' + \mu\partial_t \mathbf{H}' = -\mathbf{M}'\\
-\nabla\times \mathbf{H}' + \varepsilon\partial_t \mathbf{E}' = -\mathbf{J}'
\end{cases}
\end{equation}
From the electromagnetic fields in the wave domain, the frequency-domain fields can be computed on
the fly during modelling using the fictitious wave transformation
\begin{subequations}\label{eq:transform}
  \begin{align}
E_i'(\mathbf{x},\omega') =& \int_0^{T_{\max}} E_i'(\mathbf{x},t)e^{\mathrm{i}\omega't} \mathrm{d}t,\\
H_i'(\mathbf{x},\omega') =& \int_0^{T_{\max}} H_i'(\mathbf{x},t)e^{\mathrm{i}\omega't} \mathrm{d}t,
\end{align}
\end{subequations}
where $i=\{x,y,z\}$ are component indices of the electromagnetic fields; $T_{\max}$ is the final time until the electromagnetic fields reach their steady state.

The Green's function for point source is then obtained by normalizing the electromagnetic field with the source current. In the absence of magnetic source ($\mathbf{M}=0$),  the Green's functions due to the electrical current $J_j(\mathbf{x}_s,\omega)$ are
 \begin{subequations}\label{eq:greenfunction}
   \begin{align}
  G_{ij}^{EE}(\mathbf{x},\omega;\mathbf{x}_s) =& \frac{E_i(\mathbf{x},\omega;\mathbf{x}_s)}{J_j(\mathbf{x}_s,\omega)}
  =\sqrt{\frac{-\mathrm{i}\omega}{2\omega_0}}\frac{E_i'(\mathbf{x},\omega;\mathbf{x}_s)}{J_j'(\mathbf{x}_s,\omega)}, \\
  G_{ij}^{HE}(\mathbf{x},\omega;\mathbf{x}_s) = &\frac{H_i(\mathbf{x},\omega;\mathbf{x}_s)}{J_j(\mathbf{x}_s,\omega)}
  =\frac{H_i'(\mathbf{x},\omega;\mathbf{x}_s)}{J_j'(\mathbf{x}_s,\omega)},
  \end{align}
\end{subequations}
where  $G_{ij}^{EE}(\mathbf{x},\omega;\mathbf{x}_s)$ and $G_{ij}^{HE}(\mathbf{x},\omega;\mathbf{x}_s)$ stand for the $i$th component of the electrical and magnetic Green's function for angular frequency $\omega$ at the spatial location $\mathbf{x}$.

\section{Computer implementation}

In this section we present the key ingredients to efficiently implement the fictitious wave domain CSEM modelling based on our newly developed high-order FDTD method on non-uniform grid \citep{Yang_2023_HFDNU}. For the convenience of code implementation, we define a data structure for electromagnetic fields and use the pointer \verb|emf| to access information of all relevant parameters.

\subsection{High order FDTD over nonuniform grid}

Different from the attenuative formulation after \citep{Maao_2007_FFT}, equation \eqref{eq:mittet} is a pure wave domain equation, which can be discretized using leap-frog staggered grid finite difference method on the staggered (Yee) grid:
\begin{equation}
\begin{cases}
H_i^{'n+1/2} =& H_i^{'n-1/2} - \Delta t\mu^{-1} (\xi_{ijk}\partial_j 
E_k^{'n} + M_i^{'n})\\
 E_i^{'n+1} =&  E_i^{'n} + \Delta t (\varepsilon'^{-1})_{ij}
(\xi_{jkl}\partial_k H_l^{'n+1/2} -J_j^{'n+1/2}) 
\end{cases}
\end{equation}
where $\Delta t$ is the temporal sampling, $\xi_{ijk}$ is the Levi-Civita symbol. The positioning of each field on the nonuniform staggered grid  is similar to the staggered FDTD on a uniform grid. The major difference between FDTD implementations on a uniform and a nonuniform grids lies in the discretization of these spatial derivatives ($\partial_i E'_k$ and $\partial_k H'_l$)

In order to compute the electromagnetic field as well as its derivatives with arbitrary grid spacing, we have to do polynomial interpolation using a number of knots $x_0, x_1,\cdots, x_n$. According to Taylor expansion, we have
\begin{equation}
  f(x_i) = f(x) + f'(x)(x_i-x) + \frac{1}{2}f''(x)(x_i-x)^2 + \cdots +
  \frac{1}{n!}f^{(n)}(x)(x_i-x)^n+\cdots 
\end{equation}
where $i=0,1,\cdots,n$. Let us define $a_i(x):=f^{(i)}(x)/i!$ and use first $n+1$ distinct nodes $x_0,x_1,\cdots, x_n$ to build a matrix system
\begin{equation}
  \underbrace{  \begin{bmatrix}
      f(x_0)\\
      f(x_1)\\
      \vdots\\
      f(x_n)
  \end{bmatrix}}_{\mathbf{f}}
  =\underbrace{\begin{bmatrix}
      1 & x_0-x & (x_0-x)^2 & \cdots & (x_0-x)^n\\
      1 & x_1-x & (x_1-x)^2 & \cdots & (x_1-x)^n\\
      \vdots & \ddots & & \vdots\\
      1 & x_n-x & (x_n-x)^2 & \cdots & (x_n-x)^n\\
  \end{bmatrix}}_{\mathbf{V}^T(x_0-x,\cdots,x_n-x)} \underbrace{\begin{bmatrix}
      a_0(x)\\
      a_1(x)\\
      \vdots\\
      a_n(x)
  \end{bmatrix}}_{\mathbf{a}}
\end{equation}
where $\mathbf{V}^T(x_0-x,\cdots,x_n-x)$ is the transpose of a Vandermonde matrix determined by $x_0-x,\cdots,x_n-x$. The above expression implies that the function $f(x)$ and its derivatives up to $n$-th oder at arbitrary location $x$ can be found by inverting the Vandermonde matrix $\mathbf{a}=[\mathbf{V}^T]^{-1}\mathbf{f}$. When the order of the Vandermonde matrix increases, the inversion of Vandermonde matrix becomes highly ill-conditioned. Fortunately, there exists accurate and efficient algorithm \citep{bjorck1970solution} to achieve this challenging job, by exploring the equivalence between Vandermonde matrix inversion and the Lagrange polynomial interpolation. The following code snippet implements it following \citep[Algorithm 4.6.1]{Golub_1996_MATCOMP}.

\begin{verbatim}                                                                                    
/* solve the vandermonde system V^T(x0,x1,...,xn) a = f
 * where the n+1 points are prescribed by vector x=[x0,x1,...,xn];
 * the solution is stored in vector a.
 */
void vandermonde(int n, float *x, float *a, float *f)
{
  int i,k;

  /* calculate Newton representation of the interpolating polynomial */
  for(i=0; i<=n; ++i) a[i] = f[i];

  for(k=0; k<n; k++){
    for(i=n; i>k; i--){
      a[i] = (a[i]-a[i-1])/(x[i]-x[i-k-1]);
    }
  }

  for(k=n-1; k>=0; k--){
    for(i=k; i<n; i++){
      a[i] -= a[i+1]*x[k];
    }
  }
}
\end{verbatim}

Let the $i$-th row, $j$-th column of the inverse matrix $[\mathbf{V}^T]^{-1}$ be $w_{ij}$, i.e. $([\mathbf{V}^T]^{-1})_{ij}=w_{ij}, i,j=0,\cdots,n$. It follows that
\begin{equation}\label{eq:amatrix}
  \underbrace{  \begin{bmatrix}
      a_0(x)\\
      a_1(x)\\
      \vdots\\
      a_n(x)
  \end{bmatrix}}_{\mathbf{a}}:=\underbrace{\begin{bmatrix}
      w_{00} & w_{01} & \cdots w_{0n}\\
      w_{10} & w_{11} & \cdots w_{1n}\\
      \vdots\\
      w_{n0} & w_{n1} & \cdots w_{nn}\\
  \end{bmatrix}}_{[\mathbf{V}^T]^{-1}}\underbrace{\begin{bmatrix}
      f(x_0)\\
      f(x_1)\\
      \vdots\\
      f(x_n)
  \end{bmatrix}}_{\mathbf{f}}
\end{equation}

There are a number of situations that requires the use of interpolation weights repeatedly at every time step:
\begin{enumerate}
\item The same finite difference coefficients will be used at the same grid location to do numerical modelling;
\item The same interpolation weights may be used to record fields at the same location in different time steps.
\item The injection of the source time function needs the same interpolation weights to be applied.
\end{enumerate}
The last two cases are due to the fact that the sources and the receivers may reside in the arbitrary location, not necessarily on the grid. As a consequence, it is necessary to tabulate these weights instead of repeating the computation according to the inversion algorithm. Since applying the algorithm in operator form does not need the matrix coefficients explicitly, it is not obvious how to obtain the interpolation weights $w_{ij}$. We present an effective way here to make it. The idea is the following: feeding the algorithm with the vector $f$ with $j$th element 1 and other elements 0, the output of the Vandermonde inversion algorithm will be the $j$th column of the matrix $[\mathbf{V}^T]^{-1}$:
\begin{equation}
  \mathbf{f}=[0,\cdots,1,\cdots,0]^T, \quad \mathbf{a} = [w_{1j},\cdots,w_{nj}]^T.
\end{equation}
By looping over index $j$ from 0 to $n$, we obtain all columns of $[\mathbf{V}^{T}]^{-1}$. Over the rectilinear non-uniform grid, we construct multi-dimensional finite difference stencil using the tensor product of three 1D interpolation operator.

Consider the staggered finite difference approximation of the first derivatives in $x$ direction using $2r$ non-equidistant nodes centered at $x=x_{i+1/2}$  for the magnetic component and $x=x_i$ for the electric component. The forward operator $D_x^+$ to discretize 1st derivative of electrical field is given by
\begin{equation}\label{eq:HFD1}
  \begin{split}
    D_x^+ u |_{x=x_{i+1/2}}=& \Big(c^+_1(x_{i+1/2}) u(x_{i+1}) + c^+_{0}(x_{i+1/2}) u(x_{i})\Big) \\
    + &\Big(c^+_2(x_{i+1/2}) u(x_{i+2}) + c^+_{-1}(x_{i+1/2}) u(x_{i-1})\Big)\\
    + & \cdots \\
    + &\Big( c^+_r(x_{i+1/2}) u(x_{i+r}) + c^+_{-r+1}(x_{i+1/2}) u(x_{i-r+1})\Big)
    \end{split}
\end{equation}
and the backward operator $D_x^-$ to discretize 1st derivative of magnetic field is
\begin{equation}\label{eq:HFD2}
  \begin{split}
    D_x^- u |_{x=x_{i}}=&\Big(c^-_1(x_{i}) u(x_{i+1/2}) + c^-_0(x_{i})  u(x_{i-1/2})\Big)\\
    +&\Big(c^-_2(x_{i}) u(x_{i+3/2}) + c^-_{-1}(x_{i}) u(x_{i-3/2})\Big)\\
    +& \cdots\\
    +&\Big(c^-_r(x_{i}) u(x_{i+r-1/2}) + c^-_{-r+1}(x_{i}) u(x_{i-r+1/2})\Big),
  \end{split}
\end{equation}
where we have paired the differencing terms to highlight the similarity to the uniform grid staggered finite differencing. The coefficients $c_j^+$ and $c_j^-$, $j=-r+1,\cdots, r$ are the 2nd row of the inverse of the matrix $\mathbf{V}^T(x_{i+r-1/2}-x_i,\cdots,x_{i-r+1/2}-x_i)$ and $\mathbf{V}^T(x_{i+r}-x_i,\cdots,x_{i-r+1}-x_i)$ in \eqref{eq:amatrix}.  Using $2r$ nodes, we achieve higher accuracy up to $2r$-th order in space.

Denote the radius $r$ as \verb|emf->rd|. Using Vandermonde matrix inversion, we compute the finite difference coefficients of $2r$th order and store it in  \verb|emf->v3s| (here 3 denotes z direction and \verb|s| means grid staggering) to compute $\partial_z H_x$ and $\partial_z H_y$, yielding
\begin{verbatim}
//i1min,i1max=lower and upper bounds for i1 - x direction
//i2min,i2max=lower and upper bounds for i2 - y direction
//i3min,i3max=lower and upper bounds for i3 - z direction
for(i3=i3min; i3<=i3max; ++i3){
  for(i2=i2min; i2<=i2max; ++i2){
    for(i1=i1min; i1<=i1max; ++i1){
     ...
     if(emf->rd==1){
       D3H2 = emf->v3s[i3][0]*emf->H2[i3-1][i2][i1]
         + emf->v3s[i3][1]*emf->H2[i3][i2][i1];
       D3H1 = emf->v3s[i3][0]*emf->H1[i3-1][i2][i1]
         + emf->v3s[i3][1]*emf->H1[i3][i2][i1];
     }else if(emf->rd==2){
       D3H2 = emf->v3s[i3][0]*emf->H2[i3-2][i2][i1]
         + emf->v3s[i3][1]*emf->H2[i3-1][i2][i1]
         + emf->v3s[i3][2]*emf->H2[i3][i2][i1]
         + emf->v3s[i3][3]*emf->H2[i3+1][i2][i1];
       D3H1 = emf->v3s[i3][0]*emf->H1[i3-2][i2][i1]
         + emf->v3s[i3][1]*emf->H1[i3-1][i2][i1]
         + emf->v3s[i3][2]*emf->H1[i3][i2][i1]
         + emf->v3s[i3][3]*emf->H1[i3+1][i2][i1];
     }else if(emf->rd==3){
       D3H2 = emf->v3s[i3][0]*emf->H2[i3-3][i2][i1]
         + emf->v3s[i3][1]*emf->H2[i3-2][i2][i1]
         + emf->v3s[i3][2]*emf->H2[i3-1][i2][i1]
         + emf->v3s[i3][3]*emf->H2[i3][i2][i1]
         + emf->v3s[i3][4]*emf->H2[i3+1][i2][i1]
         + emf->v3s[i3][5]*emf->H2[i3+2][i2][i1];
       D3H1 = emf->v3s[i3][0]*emf->H1[i3-3][i2][i1]
         + emf->v3s[i3][1]*emf->H1[i3-2][i2][i1]
         + emf->v3s[i3][2]*emf->H1[i3-1][i2][i1]
         + emf->v3s[i3][3]*emf->H1[i3][i2][i1]
         + emf->v3s[i3][4]*emf->H1[i3+1][i2][i1]
         + emf->v3s[i3][5]*emf->H1[i3+2][i2][i1];
     }
     ...
    }
  }
}
\end{verbatim}

The above procedure allows us to use high order finite difference scheme to accurately compute the electromagnetic fields and their derivatives, typically with arbitrary grid spacing in the rectilinear grid. This opens the door for CSEM modelling using high order FDTD on a nonuniform grid in a consistent framework.

\subsection{Stability condition and dispersion error}

Following the standard von Neumann analysis, the author have proved \citep{Yang_2023_HFDNU} that the following condition must be satisfied in the generic rectilinear nonuniform grid for stable numerical modelling
\begin{equation}
 \eta\Delta t\leq 1 \quad\text{with} \quad
  \eta=\frac{1}{2} v_{\max}\sqrt{(D_x^{\max})^2 + (D_y^{\max})^2 + (D_z^{\max})^2},
\end{equation}
where $v_{\max}=\max\{v\}$ in which $v:=1/\sqrt{\mu\epsilon}$ is the velocity of the propagating EM field; $D_x^{\max}$, $D_y^{\max}$ and $D_z^{\max}$ are the maximum value of the the discretized first derivatives along $x$, $y$ and $z$ directions. In particular, we have
\begin{equation}
  D_x^{\max}  = \max(\sum_{i=-r+1}^r |c_i^+|,\sum_{i=-r+1}^r |c_i^-|)
\end{equation}
and similar estimations for $D_y^{\max}$ and $D_z^{\max}$. To minimize the number of time steps $N_t$, we would like to use the largest possible $\Delta t$ without violation of the above stability condition. The temporal sampling $\Delta t$ is therefore automatically determined based on the pre-determined grid spacing, i.e., $\Delta t=0.99/\eta$.

As shown in the Appendix, the dispersion error on the uniform grid can be easily measured. However, over the non-uniform grid, the discritized spatial derivative operators in \eqref{eq:HFD1} and \eqref{eq:HFD2} immediately make the dispersion analysis difficult (the wavenumbers are also space dependent). In view of the difficulties to practically measure the dispersion error on nonuniform grid, there is no control of the dispersion error in wave domain. Another solid reason for not doing so is that the final electromagnetic field we compute is in diffusion domain instead of wave domain. Physically speaking, the diffusive EM field is dispersive in itself, thus the physical dispersion will domainate the computed field after conversion from wave domain into diffusion domain. This means the numeric dispersion error in wave domain will eventually be annihilated by the physical dispersion. Note that  in our method, the physical dispersion has been achieved analytically according to equation \eqref{eq:correspondence}, thus free of numeric error.

\subsection{Top boundary condition}

At each time step, the top boundary (air-water interface in marine CSEM or air-formation interface in land CSEM) must be properly implemented with high order FDTD \citep{wang1993finite,oristaglio1984diffusion}. In the air, the conductivity is zero. Without electrical source, the Amp\`ere's law then reduces to
\begin{equation}
\nabla\times \mathbf{H} = 0,
\end{equation}
which leads to the wave-number domain relation
\begin{equation}
H_x = \frac{k_x}{k_z} H_z, \quad H_z=\frac{k_y}{k_z} H_z,
\end{equation}
where $k_x$, $k_y$ and $k_z$ are wave-numbers in x-, y- and z- directions. Since $\nabla\cdot \mathbf{H}=0$, from the relation $\nabla\times\nabla\times \mathbf{H}=\nabla(\nabla\cdot \mathbf{H})-\Delta \mathbf{H}$ we obtain
\begin{equation}
\Delta \mathbf{H} = 0,
\end{equation}
which implies $k_x^2 + k_y^2 + k_z^2=0$, hence $k_z=\pm \mathrm{i} \sqrt{k_x^2 + k_y^2}$. It is critical to choose a correct sign here such that the field vanishes at infinite far distance ($\mathbf{H}\rightarrow 0$ when $z\rightarrow 0$), yielding $k_z=-\mathrm{i} \sqrt{k_x^2 + k_y^2}$. This leads to the following implementation
\begin{equation}
H_x = \frac{\mathrm{i} k_x}{\sqrt{k_x^2+k_y^2}} H_z, \quad H_z=\frac{\mathrm{i}k_y}{\sqrt{k_x^2+k_y^2}} H_z.
\end{equation}
Note we have made a sign correction to \citep[equation 46]{Mittet_2010_HFD}. In order to use high-order FDTD scheme, we also need to extrapolate electrical fields by forcing  \citep{oristaglio1984diffusion},
\begin{equation}
\Delta \mathbf{E} = 0
\end{equation}
at the air interface.

The above boundary conditions result in the extrapolation of the fields to ghost points using Fourier transform. Because the use of fast Fourier transform (FFT) assumes equal grid spacing, the air-water boundary condition requires a uniform grid in both x- and y- directions. The use of non-uniform grid in x- and y- directions is feasible by interpolating between non-uniform grid and uniform grid. Our numerical result \citep{Yang_2023_HFDNU} shows that it leads to degraded modelling accuracy and thus not implemented in \verb|libEMM|.

\subsection{Absorbing boundary condition}

In order to mimic the wave propagation to infinity, we apply perfectly matched layer (PML) \citep{Chew_1994_PMM} boundary condition in other directions except the top part of the model. It helps to attenuate the potential reflection using truncated computing mesh. The application of PML is equivalent to perform  coordinate stretching $\partial_{\tilde{x}}=s_x^{-1}\partial_x$ with a complex factor $s_x$ ($|s_x|>1$). The convolutional PML (CPML) \citep{Roden_2000_CPML} uses $s_x = \kappa_x + \frac{d_x(x)}{\alpha_x + i\omega} (\kappa_x\ge 1)$ such that
\begin{equation}
  \partial_{\tilde{x}} = \frac{1}{\kappa_x}\partial_x +\psi_x,
\end{equation}
where the memory variable $\psi_x$ can then be computed in a recursive manner during time stepping
\begin{equation}
    \psi_x^{n+1}= b_x \psi_x^n + a_x \partial_x^{n+1/2}
\end{equation}
where
\begin{equation}
  b_x = e^{-(d_x/\kappa_x+\alpha_x)\Delta t}, a_x = \frac{d_x}{\kappa_x(d_x+\kappa_x \alpha_x)} (b_x-1). 
\end{equation}
The damping profile $d_x(x)$  and the constant $\kappa_x$ are chosen according to \citep{Komatitsch_2007_GEO}.
We implement the CPML for $x$, $y$ and $z$ directions in the same way in FDTD. Assume the resistivity/conductivity model is cube of size \verb|n1*n2*n3|. The model will be padded with \verb|nb| CPML layers and \verb|ne| number of  extra buffers on each side:\\
\begin{verbatim}
    emf->nbe = emf->nb + emf->ne;
    emf->n1pad = emf->n1 + 2*emf->nbe;//total number of gridpoints in x
    emf->n2pad = emf->n2 + 2*emf->nbe;//total number of gridpoints in y
    emf->n3pad = emf->n3 + 2*emf->nbe;//total number of gridpoints in z
\end{verbatim}
An example for the memory variables associated with $\partial_z H_x$ and $\partial_z H_y$ are given as follows:
\begin{verbatim}
    ...
    if(i3<emf->nb){
      emf->memD3H2[i3][i2][i1] = emf->b3[i3]*emf->memD3H2[i3][i2][i1] 
                               + emf->a3[i3]*D3H2;
      emf->memD3H1[i3][i2][i1] = emf->b3[i3]*emf->memD3H1[i3][i2][i1]
                               + emf->a3[i3]*D3H1;
      D3H2 += emf->memD3H2[i3][i2][i1];
      D3H1 += emf->memD3H1[i3][i2][i1];
    }else if(i3>emf->n3pad-1-emf->nb){
      j3 = emf->n3pad-1-i3;
      k3 = j3+emf->nb;
      emf->memD3H2[k3][i2][i1] = emf->b3[j3]*emf->memD3H2[k3][i2][i1] 
                               + emf->a3[j3]*D3H2;
      emf->memD3H1[k3][i2][i1] = emf->b3[j3]*emf->memD3H1[k3][i2][i1] 
                               + emf->a3[j3]*D3H1;
      D3H2 += emf->memD3H2[k3][i2][i1];
      D3H1 += emf->memD3H1[k3][i2][i1];
    }
    ...
\end{verbatim}
After this step, the curl operator of the fields can then be obtained. For the curl of magnetic fields, the formula
\begin{displaymath}
  \nabla\times \mathbf{H}=\begin{bmatrix}
  \partial_y H_z-\partial_z H_y,
  \partial_z H_x-\partial_x H_z,
  \partial_x H_y-\partial_y H_x\end{bmatrix}^T
\end{displaymath}
will translate into
\begin{verbatim}
    ...
    emf->curlH1[i3][i2][i1] = D2H3-D3H2;
    emf->curlH2[i3][i2][i1] = D3H1-D1H3;
    emf->curlH3[i3][i2][i1] = D1H2-D2H1;
    ...
\end{verbatim}
The above code implements the CPML regions and the interior regions in an elegant and compatible frame.

\subsection{Time integration}

Equation \eqref{eq:transform} shows that many time steps may be required to compute the frequency domain electromagnetic fields.  The frequency domain data should be integrated on the fly during forward modelling process thanks to the discrete time Fourier transform (DTFT).
\begin{equation}
u(\mathbf{x},\omega)=\sum_{n=0}^{N_t-1} u(\mathbf{x},t_n) \exp(\mathrm{i}\omega' t_n),
\end{equation}
where the time has been discretized as $t_n=n\Delta t$; $u\in\{E_x,E_y,E_z,H_x, H_y,H_z\}$ is one component of the electromagnetic fields; $N_t$ is the total number of time steps needed. The following code snippet examplifies the DTFT for $E_x$ component at \verb|it| time step:
\begin{verbatim}
void dtft_emf(emf_t *emf, int it, float ***E1, float _Complex ****fwd_E1)
/*<DTFT of the electromagnetic fields (emf) >*/
{
  int i1,i2,i3,ifreq;
  float _Complex factor;
  
  int i1min=emf->nb;//lower bound for index i1
  int i2min=emf->nb;//lower bound for index i2
  int i3min=emf->nb;//lower bound for index i3
  int i1max=emf->n1pad-1-emf->nb;//upper bound for index i1
  int i2max=emf->n2pad-1-emf->nb;//upper bound for index i2
  int i3max=emf->n3pad-1-emf->nb;//upper bound for index i3

  for(ifreq=0; ifreq<emf->nfreq; ++ifreq){
    factor = emf->expfactor[it][ifreq];

    for(i3=i3min; i3<i3max; ++i3){
      for(i2=i2min; i2<i2max; ++i2){
        for(i1=i1min; i1<i1max; ++i1){
          fwd_E1[ifreq][i3][i2][i1] += E1[i3][i2][i1]*factor;
        }
      }
    }
  }
}
\end{verbatim}
where \verb|E1| and \verb|fwd_E1| stand for time-domain and frequency-domain $E_x$ field. Note that we only perform the computation in the region of interest without absorbing boundaries by specifying the index bounds. We do the same for all other field components.

There have been a significant amount of effort in the literature to estimate the total modelling time $T_{\max}$ \citep{wang1993finite,Mittet_2010_HFD}. The key for this estimation is to ensure that the frequency domain EM field reaches its steady state.  In practice, the final number of time steps is usually over estimated for safety. In our implementation, we therefore regularly check the convergence of the  frequency domain field until no contribution added after more number of time steps. This can be computationally expensive if the convergence check is very frequent. This is particularly true if one wishes to output the Green's function at every grid point for every frequency of interest. Here, we take a short cut based on judicious physical intuition. Physically, the lowest frequency component of the CSEM has the largest penetration depth, according to the relation between the frequency and the skin depth $\delta=\sqrt{2/(\omega\mu\sigma)}$. Thus, using lowest frequency at the boundaries of the computing volume should be sufficient to monitor the evolution of the fields. The simulation will be automatically terminated once the field is converged, thus avoiding extra computation.

\subsection{Medium homogenization}

When an interface is present in the medium, additional effort is required to handle the high medium contrast in order to achieve precise modelling. This can be achieved by averaging the medium \citep{davydycheva2003efficient}.  \verb|libEMM| adapts it for VTI medium, even though this method is capable to handle full anisotropy. The key idea of this method is to compute the effective medium parameters for the horizontal conductivity and vertical resistivity by averaging over the cell size
\begin{subequations}
  \begin{align}
\bar{\sigma}_{xx}(x_{i+0.5}) =&\frac{1}{x_{i+1}-x_i} \int_{x_i}^{x_{i+1}}\sigma_{xx}(x)\mathrm{d}x,\\
\bar{\sigma}_{yy}(y_{j+0.5}) =&\frac{1}{y_{j+1}-y_j} \int_{y_j}^{y_{j+1}}\sigma_{yy}(y)\mathrm{d}y,\\
\bar{\sigma}_{zz}(z_{k+0.5}) =&\left(\frac{1}{z_{k+1}-x_k} \int_{z_k}^{z_{k+1}}\sigma_{zz}^{-1}(z)\mathrm{d}z\right)^{-1}.
\end{align}
\end{subequations}
The above procedure is based on the fact that the tangential field components see the model as a combination of resistors in parallel, while the normal field component sees the model as a combination of resistors in series.
In the convention of Soviet literature, the method is often coined homogenization, as 
the same formula holds for all grid points, regardless of whether the point is in a homogeneous region or in the neighbourhood of an interface.

\subsection{Optimal nonuniform grid generation}

Our finite difference modelling is carried out on a rectilinear mesh, which is simply the tensor product of 1D non-equispaced meshes. To generate these 1D meshes, the rule of geometrical progression is used \citep[Appendix C]{mulder2006multigrid}. 
Assume we have the total grid length $L$ divided into $n$ intervals ($n+1$ nodes) with a common ratio $q>1$. Denote the smallest interval $\Delta x=x_1-x_0$. Thus, the relation between $L$ and $\Delta x$ is
\begin{equation}\label{eq:optnugrid}
  L = \Delta x(1 + q + \cdots + q^{n-1})= \Delta x \frac{q^n-1}{q-1}.
\end{equation}
Due to the stability requirement and the resulting computational cost in the modelling, we are restricted to the smallest interval $\Delta x$ and a given number of nodes to discretize over certain distance. Since equation \eqref{eq:optnugrid} does not yield an explicit expression for the stretching factor $q$, the question boils down to finding the optimal $q$ from fixed  $n$, $\Delta x$ and $L$.
The relation in \eqref{eq:optnugrid} is equivalent to 
\begin{equation}\label{eq:rr}
q= \underbrace{\left(\frac{L}{\Delta x}(q-1) +1\right)^{\frac{1}{n}}}_{g(q)}
\end{equation}
which inspires us to carry out a number of fixed point iterations:
\begin{equation}\label{eq:fixed}
q^{k+1} = g(q^k), \quad k=0,1,\cdots.
\end{equation}
Since $|g'(q)|<1$ holds for all $q>1$, the scheme proposed here is guaranteed to converge. This idea has been implemented in the following code snippet.
\begin{verbatim}
float create_nugrid(int n, float len, float dx, float *x)
/*< generate power-law nonuniform grid by geometric progression >*/
{
  int i;
  float q, qq;

  if(fabs(n*dx-len)<1e-15) {
    for(i=0; i<=n; i++) x[i] = i*dx;
    return 1;
  }
  
  q = 1.1;//initial guess for the solution
  while(1){
    qq = pow(len*(q-1.)/dx + 1., 1./n);
    if(fabs(qq-q)<1e-15) break;
    q = qq;
  }
  for(x[0]=0,i=1; i<=n; i++) x[i] = (pow(q,i) - 1.)*dx/(q-1.);

  return q;
}
\end{verbatim}

\subsection{Consistent naming and memory-safe programming}

Since programming 3D CSEM modelling involves sophisticated memory allocation and initialization using pointers, \verb|libEMM| implements every computational module following the same naming convention to ensure a memory-safe code. Throughout the software, all routines with names \verb|xxxx_init()| and \verb|xxxx_clos()| serve as the constructor and the destructor. This allows the C routine \verb|xxxx.c| resembling C++ class and Fortran modules. The following details the major steps of FDTD based CSEM modelling in fictitious wave domain, as sketched in Algorithm \ref{alg:fdtd}:

\begin{itemize}

\item The pointer \verb|emf| pointing to the data structure of electromagnetic fields will be initialized by \verb|emf_init(emf)|. The relevant parameters requiring computing memory (for example, \verb|emf->rho11|, \verb|emf->rho22| and \verb|emf->rho33|) will also be allocated for reading input model in \verb|emf_init(emf)|.  Another routine named \verb|emf_close(emf)| for destroying the variables allocated before will be placed in the same source file \verb|emf_init_close.c|. 

\item Following the same convention, the pointer \verb|acqui| pointing to the data structure of survey acquisition will be initialized by \verb|acqui_init()|. This initialization allocates variables and reads the input files by different MPI processor. The routine name \verb|acqui_close()| will deallocate variables after the modelling. These are enclosed in the source file \verb|acqui_init_close.c|.

\item  The interpolation operator will be initialized by \verb|interpolation_init()| prior to modelling and destroyed by \verb|interpolation_close()| following the completion of the modelling jobs.

\item The computing model must be extended with CPML layers and some extra buffer layers. This is achieved in \verb|extend_model.c|, with initialization by \verb|extend_model_init()| and variable deallocation by \verb|extend_model_close()|.

\item The frequency domain EM fields are initialized by \verb|dtft_emf_init()| and deallocated by \verb|dtft_emf_close()|. The DTFT of EM fields will be performed at every time step by \verb|dtft_emf()| in the same source file \verb|dtft_emf.c|.

\item Air-wave boundary condition are implemented in \verb|airwave_bc.c|, initialized by \verb|airwave_bc_init()| and destroyed in \verb|airwave_bc_close()|. At each time step, the airwave treatment for electrical and magnetic fields will be carried out via \verb|airwave_bc_update_E()| and \verb|airwave_bc_update_H()|, respectively.

\item The time-domain electromagnetic fields for FDTD modelling is initialized in \verb|fdtd_init()| and destroyed in \verb|fdtd_close()|. The leap-frog time stepping at each time step will call \verb|fdtd_curlH| to compute $\nabla\times \mathbf{H}$, \verb|fdtd_update_E()| to update electrical field $\mathbf{E}$, \verb|fdtd_curlE| to compute $\nabla\times \mathbf{E}$, \verb|fdtd_update_H()| to update magnetic field $\mathbf{H}$. These routines are bundled in \verb|fdtd.c|.

\item The injection of the electromagnetic source for forward simulation, done by \verb|inject_electric_src_fwd()| and \verb|inject_magnetic_src_fwd()|, resides in \verb|inject_src_fwd.c|.

\item The convergence is checked every 100 time steps by \verb|check_convergence()|.

\item After time domain modelling, the Green's function is computed via \verb|compute_green_function()|.
  
\item The CSEM data is finally extracted via \verb|extract_emf()| and written out by \verb|write_data()|.
\end{itemize}

\begin{algorithm}[!tbh]
  \caption{FDTD-based CSEM modelling in fictitious wave domain}\label{alg:fdtd}
  \begin{algorithmic}[1]
  \State \verb|emf_init(emf);|\;
  \State \verb|acqui_init(acqui,emf);|\;
  \State \verb|sanity_check(emf);|\;
  \State \verb|interpolation_init();|\;
  \State \verb|extend_model_init(emf);|\;
  \State \verb|dtft_emf_init(emf);|\;
  \State \verb|airwave_bc_init(emf);|\;
  \State \verb|fdtd_init(emf);|\;
  \For{ $\mathtt{it=0,\cdots,N_t-1}$ }
    \State \verb|fdtd_curlH(emf, it);|\;
    \State \verb|inject_electric_src_fwd();|\;
    \State \verb|fdtd_update_E(emf, it);|\;
    \State \verb|airwave_bc_update_E(emf);|\;
    \State \verb|dtft_emf(emf, it);|\;
    \State \verb|fdtd_curlE(emf, it); |;\;
    \State \verb|inject_magnetic_src_fwd();|\;
    \State \verb|fdtd_update_H(emf, it);|\;
    \State \verb|airwave_bc_update_H(emf);|\;
    \If{$\mathtt{it\%100==0}$}
    \State  \verb|check_convergence(emf);|\;
    \State  if converged, break; \;
    \EndIf
    \EndFor
    \State \verb|compute_green_function(emf);|\;
    \State \verb|extract_emf();|\;
    \State \verb|write_data(acqui, emf);|\;
  \State \verb|fdtd_init(emf)|;\;
  \State \verb|airwave_bc_close(emf)|;
  \State \verb|dtft_emf_close(emf)|;\;
  \State \verb|extend_model_close(emf)|;\;
  \State \verb|interpolation_close()|;\;
  \State \verb|emf_close(emf)|;\;
  \State \verb|acqui_close(acqui)|;\;
  \end{algorithmic}
\end{algorithm}

To allocate memory for the variables used in CSEM modelling, we dynamically allocate arrays of pointers of arrays of pointers ... using contiguous memory chunk, to construct multidimensional arrays such that these arrays can be referenced in the same way as static arrays in C. That is, the $(i, j, k)$-th element can be referenced simply by \verb|array[k][j][i]|. The rightmost index \verb|i| is the fastest index while the leftmost index \verb|k| is the slowest one.  The array of size \verb|n1*n2*n3| starts from the first element \verb|array[0][0][0]|, and ends with the last one \verb|array[n3-1][n2-1][n1-1]|.

\section{Multithread parallelization on GPU}

\subsection{Strategies for GPU modelling}

The GPU acceleration technology has been mature after more than one decade of the development effort. Besides the successful applications in seismic community such as reverse time migration and full waveform inversion \citep{Yang_2014_RTM,Yang_2015_GPU}, the use of GPU for 3D CSEM is scarcely reported. Here we highlight some key strategies when implementing GPU-based 3D CSEM modeling.

\begin{figure}[!htb]
  \centering
  \includegraphics[width=0.75\textwidth]{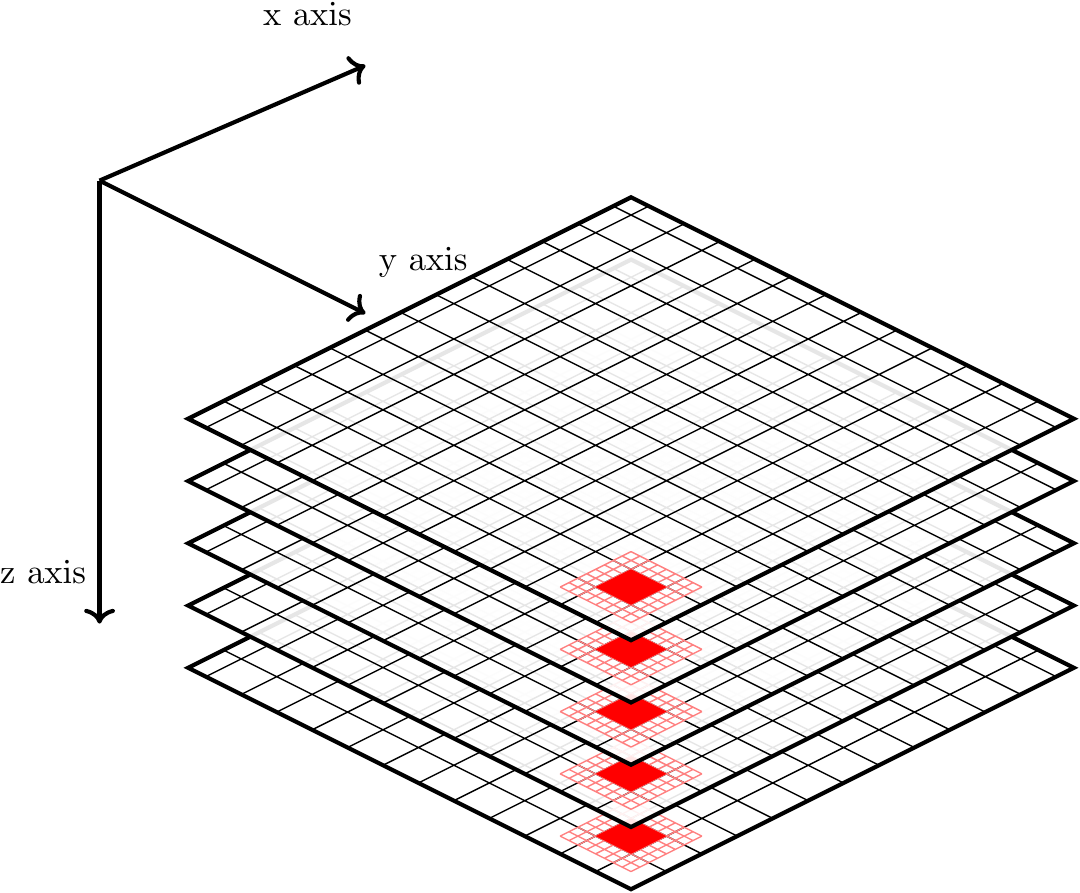}
  \caption{The tiling blocks on x-y plan along z dimension}\label{fig:diagram}
\end{figure}

\begin{figure}[!htb]
  \centering
  \includegraphics[width=0.9\textwidth]{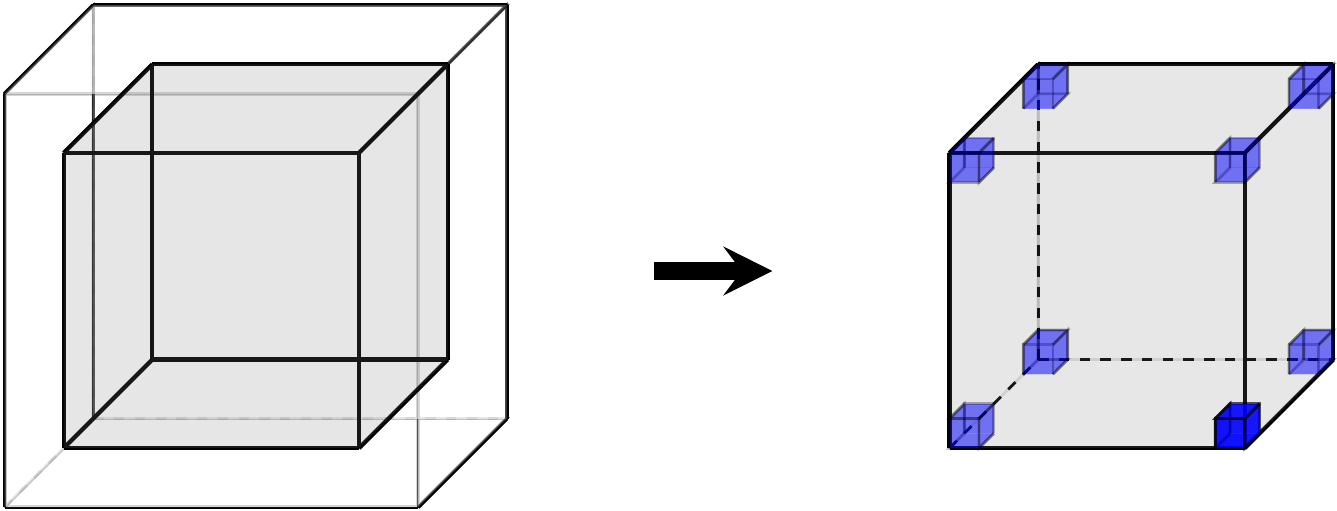}
  \caption{Schematic plot of convergence check using eight corners of the interior computing cube without absorbing boundaries}\label{fig:corners}
\end{figure}

\begin{enumerate}

  \item Perform computation intensive part on GPU without frequent CPU-GPU data traffic. Note that  CPU emphasis on low latency, while GPU emphasis on high throughput. In order to avoid substantial amount of communication latency, we design the code to perform all necessary pre-computation as much as possible before porting to GPU-based time stepping. The pre-computation includes validation of the input parameter for sanity check, preparation of the medium on extended domain, tabulating the interpolating weights at the source and receiver locations in a lookup table. 
 The air-water boundary condition is directly computed on device with CUFFT library.

 \item Use shared memory with tiled algorithm \citep{micikevicius20093d}.
 The key idea is to explore the fast L1 cache of the shared memory, which is often quite small memory size but 2 order of magnitude faster than accessing the global memory. Tiling forces multiple threads to jointly focus on a subset of the input data at each phase of execution.
 Here we map the global memory onto 2D shared memory blocks and slide it along the third dimension, as illustrated in Figure~\ref{fig:diagram}.
 The reading and writing of the data on GPU will be performed block by block manner rather than element by element, thus much more efficient. 
 Since we have to repeatedly reuse the same quantity at different location in the finite difference stencil, the tiling technique with shared memory enhances locality of data access.

\item A linear increasing index with consecutive memory access.  Each CUDA warp contains 32 processing cores executed in a single instruction multiple thread fashion. The CUDA kernel will serialize the operations when the elements reside in different warps, which leads to dramatic performance penalty. The linear increasing index with consecutive memory access is therefore crucial to design efficient the CUDA kernels.

\item Improved scheme for convergence check of the fields.
The modelling can be terminated only if the frequency domain EM fields converges at every grid point. This naturally leads to a reduction operation at all steps of convergence check. To do this efficiently on GPU, the author implemented a parallel reduction scheme in \citep{Yang_2021_GPU_CSEM}. In this paper, we do it in a more judicious manner by only checking 8 corners of interior computing cube without absorbing boundaries, as sketched in Figure~\ref{fig:corners}. This eliminates the need for parallel reduction and dramatically reduces the required number of floating point operations. The scheme can equally be applied in the sequential implementation.

\item Use of mapped pinned memory. Due to convergence check, timestepping modelling involves the communication between the device (GPU) and host (CPU), which violates the first principle aforementioned. To avoid this, we invoke mapped pinned memory which is  desirable for the CPU and GPU to share a buffer without explicit buffer allocations and copies. It is well-known that extensive use of mapped pinned memory may hit the performance. Here, only 8 corners of the complex-valued EM field has to be backed up using zero-copy pinned memory, which minimizes the potential performance deterioration.

\end{enumerate}

\subsection{CUDA implementation}

The above techniques imply that parallel GPU implementation requires a significant adaptation of each subroutine in Algorithm \ref{alg:fdtd} into different CUDA kernel functions, while the computational workflow remains the same but runs in on GPU device instead of CPU host. In GPU mode, the device must be initialized first, and then performing dedicated operations by these CUDA kernel functions, using different number of multithreaded block and grid sizes. This has been done by \verb|cuda_modelling.cu|, as a replacement of the sequential modelling conducted by \verb|do_modeling.c|.
Some simple calculations may be sophisticated to achieve but can easily performed on CPU host only once. These operations, including the extension of the model for absorbing boundaries, the setup of CPML layers and DTFD coefficients, the computation of interpolation weights to inject the source and extract data at receiver loations, have been carried out priori to \verb|cuda_modeling.cu|: GPU will directly copy them into device memory for direct utilization. This highlights the importance of exploring the distinct advantages of CPU and GPU computation.

All relevant CUDA kernels are placed in \verb|cuda_fdtd.cuh|. In particular, the computation of the spatial derivatives in FDTD are $\nabla\times \mathbf{E}$ and $\nabla\times\mathbf{H}$, which have been parallelized by GPU kernels \verb|cuda_fdtd_curlE_nugrid| and \verb|cuda_fdtd_curlH_nugrid|, as the alternatives to CPU routines \verb|fdtd_curlE| and \verb|fdtd_curlH|. According to tiled algorithm, the first two dimenions are parallelized using shared memory, while the 3rd dimension strides sequentially, yielding the following code
\begin{verbatim}
__global__ void cuda_fdtd_curlH_nugrid(...)		
{
  const int i1 = threadIdx.x + blockDim.x*blockIdx.x;
  const int i2 = threadIdx.y + blockDim.y*blockIdx.y;
  const int t1 = threadIdx.x + 2;//t1>=2 && t1<BlockSize1+2
  const int t2 = threadIdx.y + 2;//t2>=2 && t2<BlockSize2+2

  __shared__ float sH1[BlockSize2+3][BlockSize1+3];
  __shared__ float sH2[BlockSize2+3][BlockSize1+3];
  __shared__ float sH3[BlockSize2+3][BlockSize1+3];

  int in_idx = i1+n1pad*i2;
  int out_idx = 0;
  int stride = n1pad*n2pad;

  for(i3=i3min; i3<=i3max; i3++){//i3-2>=0 && i3+1<n3pad
    ...
    __syncthreads();

    if(validrw){
      ...
      curlH1[out_idx] = D2H3 - D3H2;
      curlH2[out_idx] = D3H1 - D1H3;
      curlH3[out_idx] = D1H2 - D2H1;
    }
  }//end for i3
}
\end{verbatim}
Note that it is important to synchronize GPU threads before proceeding to next stage. For efficiency, the timestepping of $\mathbf{E}$ and $\mathbf{H}$ fields also conforms the tiled mapping, i.e.,
\begin{verbatim}
__global__ void cuda_fdtd_update_E(...)
{
  const int i1 = threadIdx.x + blockDim.x*blockIdx.x;
  const int i2 = threadIdx.y + blockDim.y*blockIdx.y;
  int i3, id;
  
  for(i3=0; i3<n3pad; i3++){
    if(i1<n1pad && i2<n2pad){
      id = i1+n1pad*(i2+ n2pad*i3);
      E1[id] += dt*inveps11[id]*curlH1[id];
      E2[id] += dt*inveps22[id]*curlH2[id];
      E3[id] += dt*inveps33[id]*curlH3[id];
    }
  }
}
\end{verbatim}
The convergence of the field is checked directly on GPU by \verb|cuda_check_convergence|. Since the result of convergence check must be visible on host, we use \verb|cudaHostAlloc| instead of commonly used routine \verb|cudaMalloc|, to allocate memory to record the number of converged corners in 3D cube. Correspondingly, the pinned memory must be freed using \verb|cudaFreeHost| rather than \verb|cudaFree|. To be concrete, t, these correpond to the following lines of code:
\begin{verbatim}
  cudaHostAlloc(&h_ncorner, sizeof(int), cudaHostAllocMapped);
\end{verbatim}
and
\begin{verbatim}
  cudaFreeHost(h_ncorner);
\end{verbatim}

\section{Software features}

\subsection{Dependencies for building executable}

\verb|libEMM| is a light-weight yet powerful 3D CSEM modelling toolkit which can run using single CPU processor, multiple MPI parallel process and also CUDA-enabled GPU mode. This implies that the software has the following dependencies:
\begin{itemize}
\item FFTW for fast Fourier transform;
\item MPICH (or other types of MPI) for multi-processor parallelization over distributed memory architecture;
\item Nvidia CUDA package if one wish to benefit from GPU parallelization on device shared memory.
\end{itemize}
The installation of CUDA package is optional while the first two are mandatory. \verb|libEMM| uses the C programming language to achieve high performance computing. To compile the code, one will use the command \verb|make| to compile the code following the given \verb|Makefile|, which may be edited according to the path of the user's installation for the above packages. The compiled executable will be named as \verb|fdtd| and placed in the folder \verb|bin|.

\subsection{Parameter parsing}

\verb|libEMM| is capable to automatically parse the argument list based on the parameter specified in the token. The value for each keyword will be picked up after a equal sign \verb|=|. Multiple values for the same keyword can be specified via comma-separated parameters. Different keywords will be distinguished via the space.

\subsection{Survey acquisitions}

To specify the survey layout, three acquisition files in ASCII format must be given to prescribe the source and receiver locations as well as the their connection table. These files will be assigned to the arguments as
\begin{verbatim}
fsrc=sources.txt frec=receivers.txt fsrcrec=src_rec_table.txt
\end{verbatim}
where the file \verb|sources.txt| gives the source locations; the file \verb|receivers.txt| gives the receiver locations; the file \verb|src_rec_table.txt| establishes the connection between sources and receivers. When the code runs with MPI, each process will read these files and then carry out independent modelling jobs according to their process index (which uniquely determines a source). The source/receiver file is in the following form:
\begin{verbatim}
      x     y      z       azimuth    dip        iTx/iRx
-8000.0   0.0    275.0        0        0           1
-7975.0   0.0    275.0        0        0           2
-7950.0   0.0    275.0        0        0           3
-7925.0   0.0    275.0        0        0           4
    ...
\end{verbatim}
The above 6 columns corresponds to $(x,y,z)$ coordinates (in meters), the azimuth and dip angles in rad and the index of the source/receivers. Thanks to the interpolation operator computed by inverting Vandermonde matrix, the sources and receivers can be accurately injected/extracted at arbitrary locations without the need to sit on grid. The connection table reads
\begin{verbatim}
 iTx iRx
 1    1
 1    2
 1    3
 ...
 
 2    1
 2    2
 2    3
 ...
\end{verbatim}
which connects each source/transmitter indexed by \verb|iTx| are associated with a number of receivers index by \verb|iRx|.

\subsection{Physical domain,  grid dimensions and coordinates}

\verb|libEMM| carries out CSEM modelling over a 3D cube with physical dimensions specified by the bounds. Modelling over a domain $X=X_1\times X_2\times X_3$ with $X_1=X_2=[-10000,10000]$m, $X_3=[0,3000]$ reads
\begin{verbatim}
x1min=-10000 x1max=10000 x2min=-10000 x2max=10000 x3min=0 x3max=3000
\end{verbatim}

The parameters  \verb|n1|, \verb|n2| and \verb|n3| specify the dimension of the model in x, y and z directions, while \verb|d1|, \verb|d2| and \verb|d3| specify the \emph{smallest} grid spacing (in meters) along each coordinate. To do modelling in a resistivity cube of $101^3$ grid points, $\Delta x=\Delta y=200$ m, $\Delta z=50$ m on uniform grid, one has to write in the argument list
\begin{verbatim}
n1=101 n2=101 n3=101 d1=200 d2=200 d3=50
\end{verbatim}
These parameters will be checked by \verb|libEMM| in order to match the bounds of the domain.

Because \verb|libEMM| allows the CSEM modelling using both uniform grid and non-uniform grid, the grid coordinate in binary format, for argument \verb|fx3nu| corresponding to the gridding in z direction, must be provided in order to model on non-uniform grid along z. Note that the file must store exactly \verb|n3| floating points. Non-uniform gridding in x- and y- directions has been forbidden due to degraded modelling accuracy \citep{Yang_2023_HFDNU}. When the modelling is performed on uniform grid, there is no need to provide input for \verb|fx3nu|.

\subsection{Resistivity files}

Three resistivity files in binary format are expected to feed the argument list for numerical modelling, as specified in the following
\begin{verbatim}
frho11=rho11 frho22=rho22 frho33=rho33
\end{verbatim}
where the files \verb|rho11|, \verb|rho22| and \verb|rho33| are the resistivities of size \verb|n1*n2*n3| after homogenization, implying a half grid shift in x, y and z directions, respectively. These files should be built using model building tools outside the modelling kernel.

\subsection{Source and receiver channels, frequencies}

\verb|libEMM| specifies the active source channel and receiver channels via the keywords - \verb|chsrc| and \verb|chrec|. For example, recording CSEM data of components $E_x$ and $E_y$ from an electrical source $J_x$ reads
\begin{verbatim}
chsrc=Ex chrec=Ex,Ey
\end{verbatim}
Similarly, one can specify the simulation frequencies as
\begin{verbatim}
freqs=0.25,0.75,1.25
\end{verbatim}
in which 0.25 Hz, 0.75 Hz and 1.25 Hz will be the frequencies extracted from time-domain CSEM modelling engine.

\subsection{Operator length, CPML layers and buffers}

The half length of the finite-difference operator is determined by parameter \verb|rd|. It is possible to choose \verb|rd|=1, 2 or 3 in CPU mode, but only \verb|rd=2| is supported on GPU. This is because dramatic accuracy improvement has been observed in modelling when switching from the 2nd to the 4th order scheme, but increasing from 4th order to even higher order requires more computational effort without significant accuracy gain \citep{Yang_2023_HFDNU}.

We pad the model with the same  number of CPML layers on each side of the cube, specified by the parameter \verb|nb|. In addition, we add \verb|ne| number of buffers as a smooth transition zone between CPML and interior domain. A general principle for choosing \verb|ne| is to ensure the number of buffer layers larger than the half length of finite-difference stencil, for example, $\mathtt{ne=rd+5}$.

\subsection{CSEM modelling output}

The outcome after running a 3D CSEM modelling will be output as ASCII files named as \verb|emf_XXXX.txt|, where \verb|XXXX| is the index of the sources/transmitters. Running the simulation using 4 MPI process leads to
\begin{verbatim}
emf_0001.txt emf_0002.txt emf_0003.txt emf_0004.txt
\end{verbatim}
Each file will follow the same convention to print out the transmitter index \verb|iTx|, the receiver index \verb|iRx|, the receiver channel \verb|Ex/Ey/Ez/Hx/Hy/Hz|, the frequency index \verb|ifreq| and the real and imaginary part of the complex-valed EM field in frequency domain:
\begin{verbatim}
iTx  iRx  chrec  ifreq 	  emf_real 	   emf_imag
1    1 	   Ex 	   1 	 -1.086630e-14 	 2.026699e-16
1    2 	   Ex 	   1 	 -1.148639e-14 	 3.976523e-16
1    3 	   Ex 	   1 	 -1.213745e-14 	 6.254443e-16
1    4 	   Ex 	   1 	 -1.282039e-14 	 8.894743e-16
1    5 	   Ex 	   1 	 -1.353619e-14 	 1.192894e-15
1    6 	   Ex 	   1 	 -1.428555e-14 	 1.539961e-15
1    7 	   Ex 	   1 	 -1.506933e-14 	 1.934350e-15
1    8 	   Ex 	   1 	 -1.588794e-14 	 2.380965e-15
1    9 	   Ex 	   1 	 -1.674208e-14 	 2.884079e-15
1    10    Ex 	   1 	 -1.763176e-14 	 3.449250e-15
1    11    Ex 	   1 	 -1.855739e-14 	 4.081327e-15
...
\end{verbatim}

\subsection{Running in different modes}

To run the 3D CSEM modelling, one may write the parameters in a shell script \verb|run.sh| and launch the code using \verb|bash run.sh|. The following is an example shell script running with 25 MPI process, each of the MPI process parallelized by 4 OpenMP threads:
\begin{verbatim}
#!/usr/bin/bash

n1=101
n2=101
n3=101
d1=200
d2=200
d3=50 

export OMP_NUM_THREADS=4
mpirun -n 25 ../bin/fdtd \
       mode=0 \
       fsrc=sources.txt \
       frec=receivers.txt \
       fsrcrec=src_rec_table.txt \
       frho11=rho11 \
       frho22=rho22 \
       frho33=rho33 \
       chsrc=Ex \
       chrec=Ex \
       x1min=-10000 x1max=10000 \
       x2min=-10000 x2max=10000 \
       x3min=0 x3max=5000 \
       n1=$n1 n2=$n2 n3=$n3 \
       d1=$d1 d2=$d2 d3=$d3 \
       nb=12 ne=6 \
       freqs=0.25,0.75,1.25 \
       rd=2 
\end{verbatim}
where there are 25 sources in the acquisition. It is also possible to do modelling for some specific sources, for example, 
\begin{verbatim}
mpirun -n 2 ../bin/fdtd  shots=12,14 ...
\end{verbatim}
will only simulate CSEM data for transmitter-12 and transmitter-14. The output CSEM response will be named as \verb|emf_0012.txt| and \verb|emf_0014.txt|. The code will terminate if the number of process launched for \verb|mpirun| exceeds the total number of sources given in the file \verb|sources.txt|.

The code can be compiled with \verb|nvcc| compiler for GPU modelling using command `\verb|make GPU=1|'. 
Assume there is only one GPU card available on the laptop. Once the executable is created after compilation, one can directly run the code by executing the command with proper parameters:
\begin{verbatim}
../bin/fdtd ...
\end{verbatim}
Adding \verb|nvprof| in the script provides us a convenient way to profile the computing time spent on different CUDA kernels
\begin{verbatim}
nvprof --log-file profiling.txt ../bin/fdtd ...
\end{verbatim}
where the profiling log will be recorded in \verb|profiling.txt|.

\section{Application examples}

The software allows modelling on both uniform grid (using the default flag \verb|emf->nugrid=0|) and non-uniform grid (\verb|emf->nugrid=1|). We start with a layered model possessing a 1D structure, for which the use of uniform grid is sufficient. The high-order FDTD over non-uniform grid has been developed for practical applications to model 3D CSEM response where the part of the computational domain requires dense sampling. The second example including a varying seafloor bathymetry therefore serves as an illustration. Three frequencies, i.e., 0.25 Hz, 0.75 Hz and 1.25 Hz which are representative in real applications, are examined in this experiment. The temporal interval and the number of time steps are automatically chosen \citep{Mittet_2010_HFD} without breaking the stability condition. The relative amplitude error is measured by $|E_x^{FD}|/|E_x^{ref}|-1$. It should be close to 0 if the modelling is precise, positive if $|E_x^{FD}|>|E_x^{ref}|$ and negative if positive if $|E_x^{FD}|<|E_x^{ref}|$. The phase error is measured by $\angle E_x^{FD}-\angle E_x^{ref}$ in degrees.

\subsection{Layered model}

\begin{figure}[!htb]
  \centering
  \includegraphics[width=0.7\textwidth]{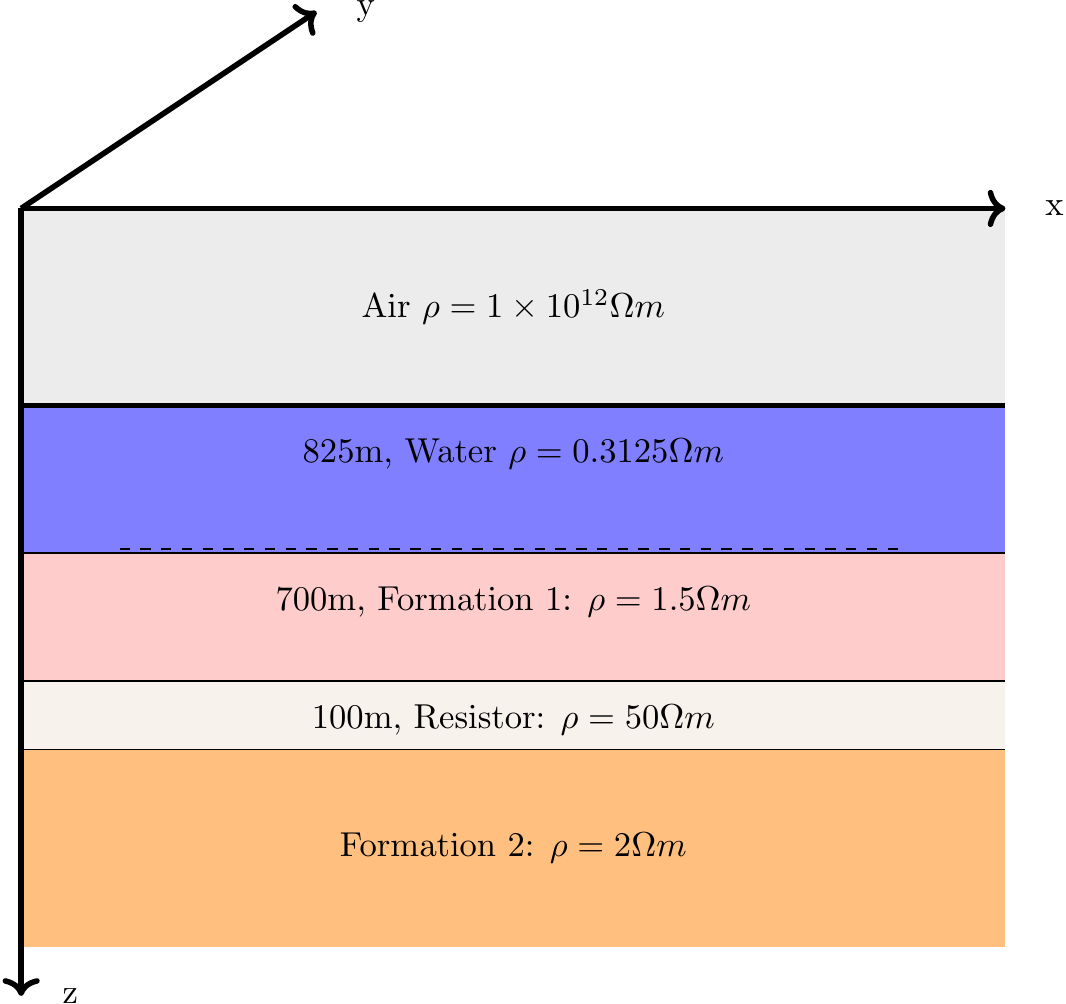}
  \caption{The cross-section of the resistivity model: the dash line indicates the depth of the sources.}\label{fig:model1d}
\end{figure}

\begin{figure}[!htb]
  \centering
  \includegraphics[width=\textwidth]{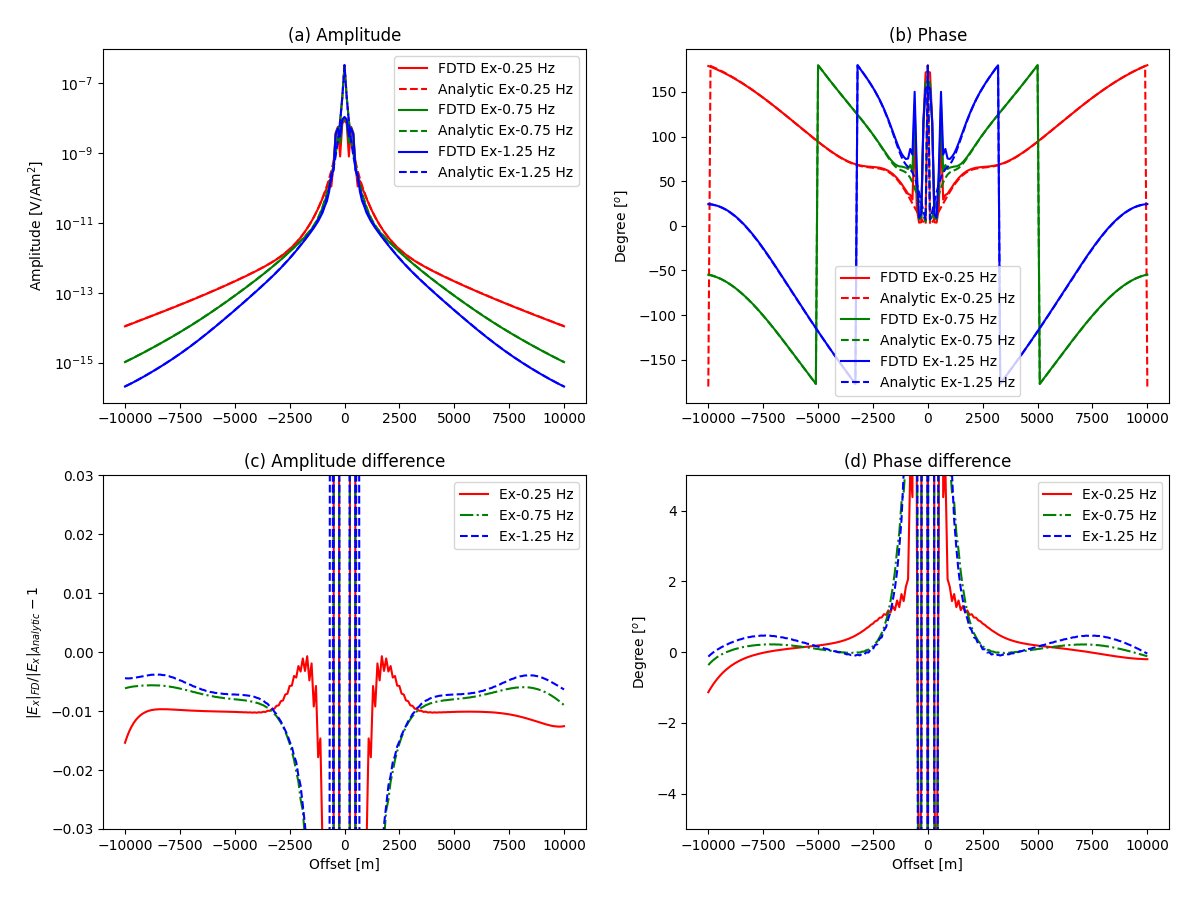}
  \caption{Comparison  between FDTD and analytic solution for 3D CSEM simulation in the shallow water scenario. (a) Amplitude; (b) Phase; (c) Amplitude error; (d) Phase error.}\label{fig:comparison}
\end{figure}

The first test is a layered resistivity model shown in Figure~\ref{fig:model1d}: the model has 5 layers, the air of $10^{12} \; \Omega\cdot \mbox{m}$, the water of $0.3125 \; \Omega\cdot \mbox{m}$, and two layers of formations. Between two layers of formations, there exists a resistive layer of $100 \; \Omega\cdot \mbox{m}$ mimicking hydrocarbon bearing formations. The 3D FDTD code is now cross-validated against the semi-analytic solution calculated by Dipole1D program \citep{key20091d}. We discretize the model of dimension $10 \; \mbox{km} \times 10\; \mbox{km} \times 5 \;\mbox{km}$ including a resistor at the depth 1250-1350 m, with grid spacing $\Delta x=\Delta y=100 \; \mbox{m}$, $\Delta z=50 \; \mbox{m}$. This results in a large 3D grid of size $101^3$. A point source is placed at 50 m above the seabed.  Figures~\ref{fig:comparison}a and \ref{fig:comparison}b show very good agreement between the FDTD method and the semi-analytic solution, in terms of the amplitude and the phase of the $E_x$ component. Figures~\ref{fig:comparison}c and \ref{fig:comparison}d show less than $1.5\%$ of the amplitude error and less than $1^\circ$ of the phase error at most of the offsets. The error at the near offset are very large. Fortunately the near offset data are not relevant for practical CSEM imaging. This example confirms the good accuracy of the FDTD method.

\subsection{Model with seafloor bathymetry}

\begin{figure}[!htb]
  \centering
  \includegraphics[width=\linewidth]{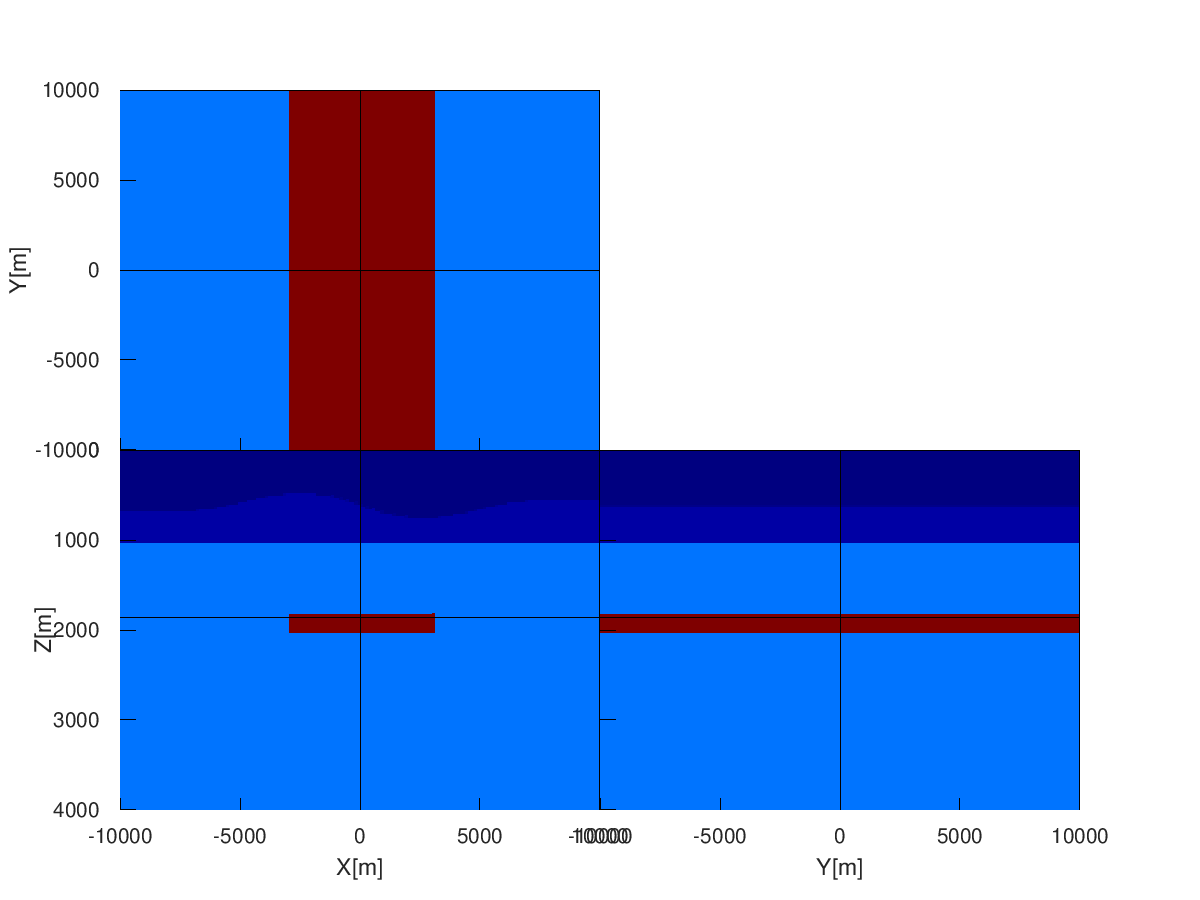}
  \caption{Marine CSEM conductivity model with seafloor bathymetry}\label{fig:mcsemmodel}
\end{figure}

\begin{figure}[!htb]
  \centering
  \includegraphics[width=\linewidth]{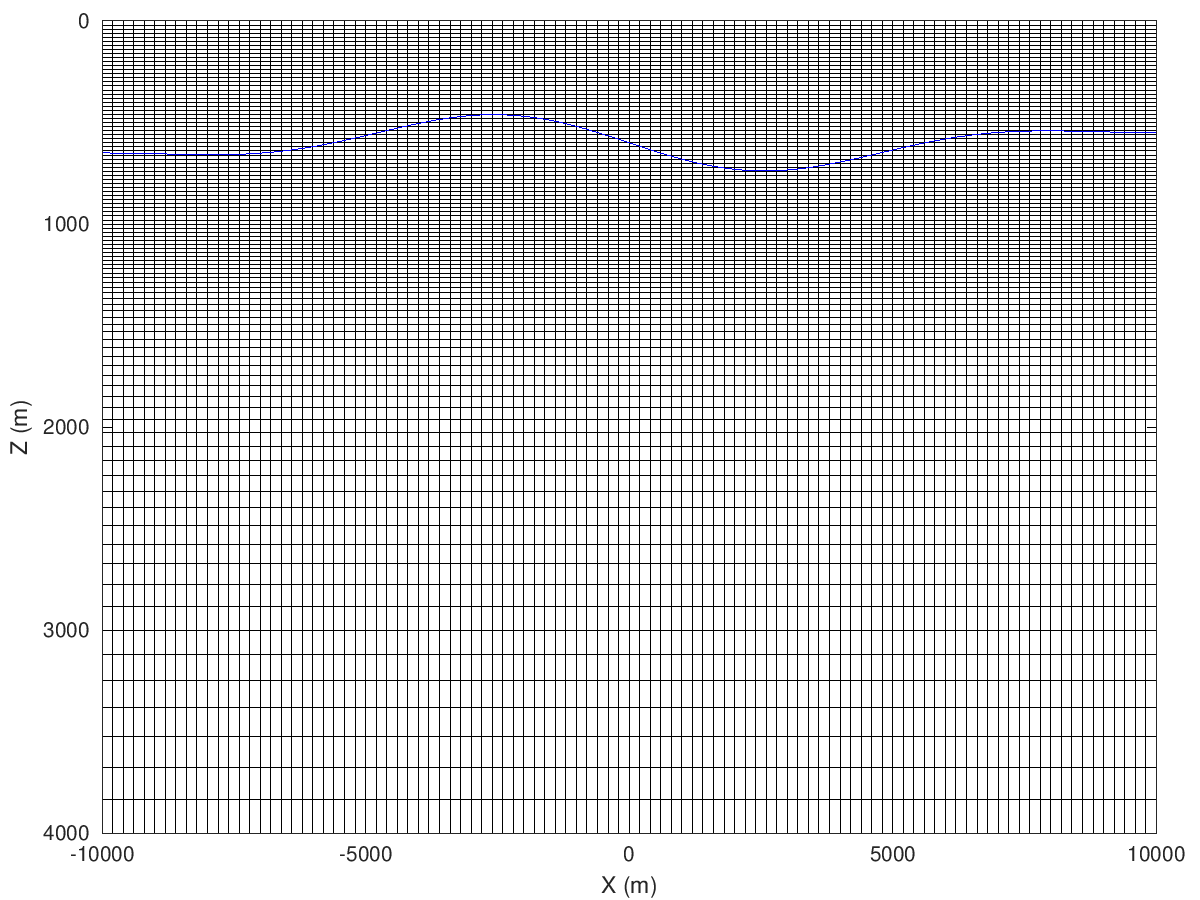}
  \caption{The non-uniform grid to mesh the resistivity including bathymetry}\label{fig:mesh}
\end{figure}

\begin{figure}[!htb]
  \centering
  \includegraphics[width=\linewidth]{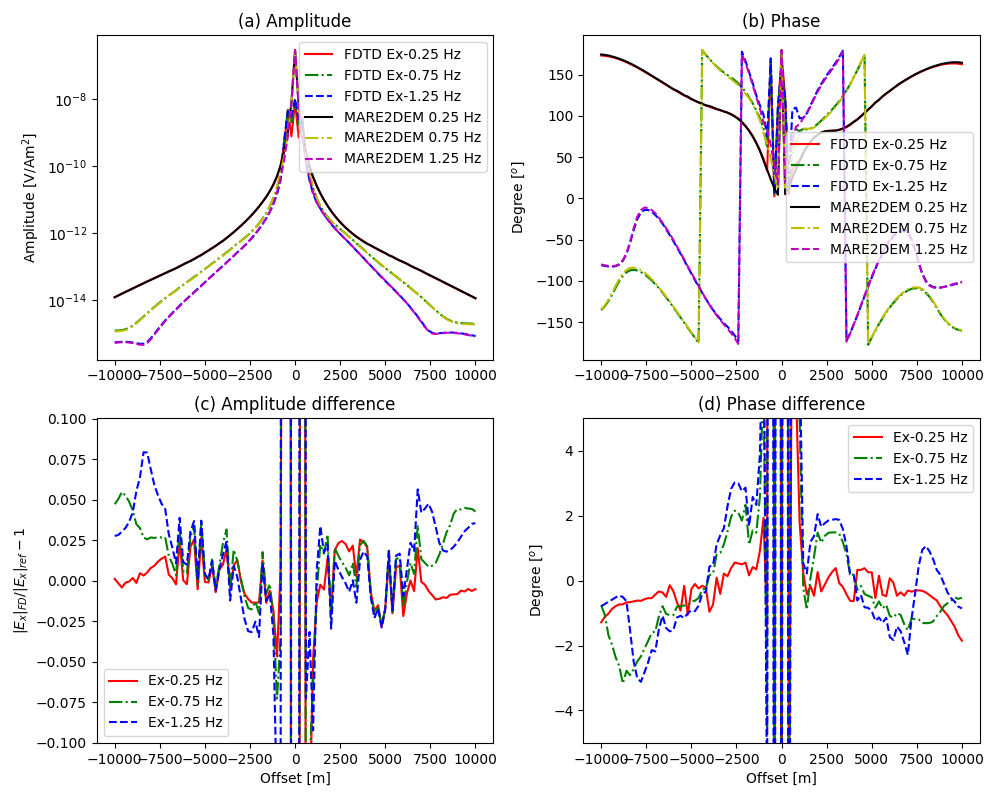}
  \caption{Comparison of simulated CSEM response by FDTD and MARE2DEM}\label{fig:bathycomparison}
\end{figure}

To mimic practical marine CSEM environment, we perform CSEM model using a 3D resistivity model of 2D structure including seafloor bathymetry, as shown in Figure~\ref{fig:mcsemmodel}. To catch the impact of seafloor bathymetry, we mesh the model densely around the seabed, while gradually stretching the grid along the depth below. The x-z section of the mesh is shown in Figure~\ref{fig:mesh}. With such a model, we compute a reference solution using MARE2DEM \citep{key2016mare2dem} using finite element method. 
 It extends the model tens of kilometers in each direction, to mimic that EM fields propagate to very far distance while avoiding possible edge/boundary effect. The regions of interest has been densely gridded using Delaunay triangulation, while big cells are employed at far distance away from interior part of the model. In our finite difference modelling, the PML boundary condition attenuates the artificial reflections in the computational domain within tens of grids to achieve the same behavior. Both the amplitude and phase match very well between FDTD modelling and MARE2DEM solution, as can be seen in Figure~\ref{fig:bathycomparison}. Figure~\ref{fig:bathycomparison}c shows that the amplitude error are bounded within 3\% for all frequencies at relevant offsets; the amplitude error at far offset and higher frequencies becomes larger when approaching to noise level ($10^{-15} \;\mbox{V/m}^2 $).
 
\begin{figure}[!htb]
  \centering
  \includegraphics[width=0.8\textwidth]{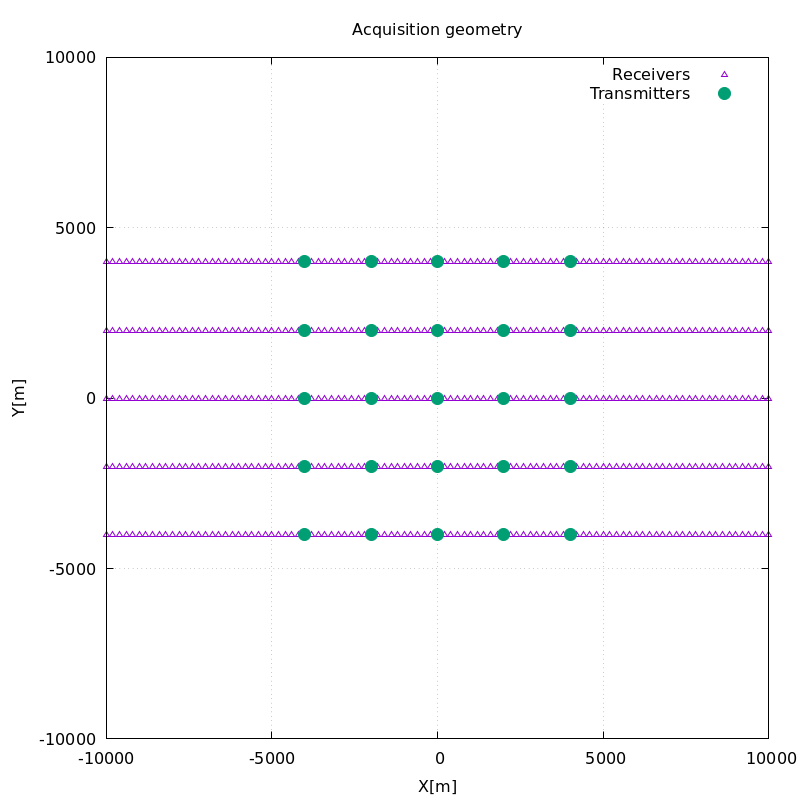}
  \caption{A survey layout sheet}\label{fig:surveylayout}
\end{figure}

\begin{figure}[!htb]
  \centering
  \includegraphics[width=\textwidth]{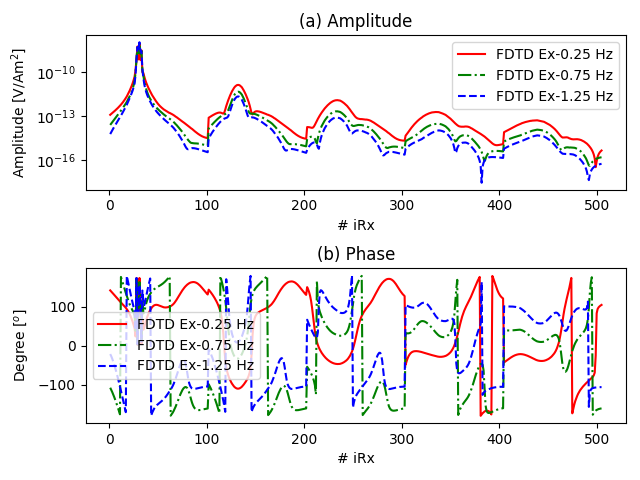}
  \caption{The amplitude and phase of the CSEM data (including inline and broadside data) modelled with a transmitter deployed at (x,y,z)=(-4000 m, -4000 m, 550 m).}\label{fig:broadside}
\end{figure}

When \verb|libEMM| is launched using multiple MPI process, we can simultaneously model CSEM data associated with different transmitters while all receivers are active/alive. This mimics the practical situations in real marine CSEM acquisitions. For a survey layout in Figure~\ref{fig:surveylayout}, \verb|libEMM| helps us to obtain both inline and broadside data, as shown in Figure~\ref{fig:broadside}.

\section{Performance analysis: OpenMP versus GPU}

Using the previous layered resistivity model of size $101^3$, we analyze the computational efficiency by comparing the runtime of the same simulation using GPU and CPU. Our GPU card is Nvidia Geforce GTX860M, which is of compute capability 5.0 paired with 2048 MB GDDR5 memory, operating at a frequency of 797 MHz. 
We compare it against Intel(R) Core(TM) i7-4710HQ CPU @ 2.50GHz.  The C code has been highly optimized and parallelized using different number of OpenMP threads.

When running in GPU mode, the profiling information recorded by CPU clock is misleading.
The command \verb|nvprof| is then useful to profile the runtime of each GPU kernel with good time resolution. The CUDA events are used as synchronization markers to accurately measure whole timing. 

\begin{figure}[!htb]
  \centering
  \includegraphics[width=\linewidth]{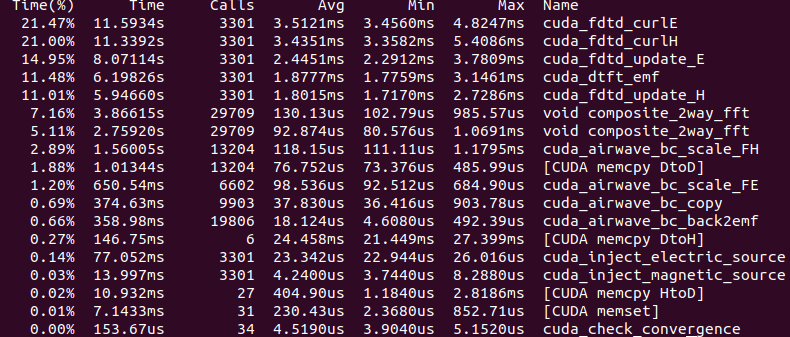}
  \caption{Screenshot of profiling log on GPU activities}\label{fig:screenshot}
\end{figure}

\begin{figure}[!htb]
  \centering
  \includegraphics[width=0.9\linewidth]{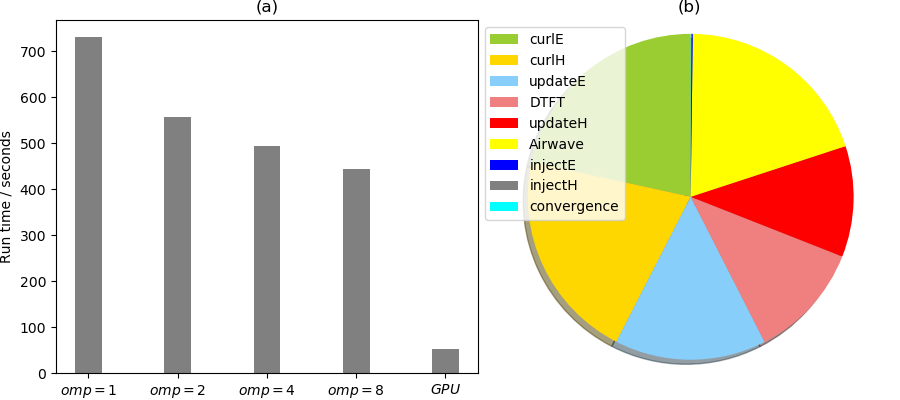}
  \caption{(a) Total runtime comparison between CPU with different number of OpenMP threads and GPU; (b) The percentage of each computing stage on GPU}\label{fig:runtime}
\end{figure}

Figure~\ref{fig:screenshot} is a screenshot of profiling log from different GPU kernels.   Table~\ref{table:cputime} summarizes the CPU runtime of the computation at all phases during different tests. Due to the existence of high resistive layer, the time-domain modelling was forced to use very small temporal sampling. The 3D modelling completes after more than 3000 time steps. It takes only approximately 53 seconds on GPU. To better catch the speedup gained by GPU compared with multithreaded CPU, we display the total runtime in Figure~\ref{fig:runtime}a.  The GPU code obtains a speedup factor of 13.7 over single threaded CPU, and a factor of 8.3 compared with CPU parallelized by 8 OpenMP threads.  We see that the performance of the code on CPU does not scale linearly with increasing number of OpenMP threads. Figure~\ref{fig:runtime}b gives a pie plot showing the proportion of each computational phase in the global GPU modelling process. The most computational intensive operations are the calculation of $\nabla\times \mathbf{E}$ and $\nabla\times\mathbf{H}$, the update of the fields $\mathbf{E}$ and $\mathbf{H}$, the application of air-wave boundary condition and the DTFT. The time spent on injecting source and convergence check is negligible.

\begin{table} 
\centering
  \caption{Runtime (in seconds) using CPU with different number of OpenMP threads}\label{table:cputime} 
  \begin{tabular}{c|c|c|c|c}   
  \hline
  Computation    & omp=1 	& omp=2       & omp=4       & omp=8 \\
  \hline  
  $\nabla\times\mathbf{E}$   	 &2.07e+02  &1.39e+02 &1.05e+02 &8.56e+01\\
  inject $\mathbf{H}$ 	 &1.12e-03  &1.73e-03 &1.48e-03 &2.10e-03\\
  udpate $\mathbf{H}$ 	 &4.18e+01  &3.36e+01 &2.91e+01 &2.66e+01\\
  $\nabla\times\mathbf{H}$   	 &1.88e+02  &1.22e+02 &9.21e+01 &7.65e+01\\
  inject $\mathbf{E}$ 	 &3.94e-03  &5.28e-03 &4.72e-03 &6.07e-03\\
  udpate $\mathbf{E}$ 	 &5.49e+01  &4.23e+01 &3.66e+01 &3.41e+01\\
  airwave 	 &2.06e+02  &1.91e+02 &2.03e+02 &1.93e+02\\
  DTFT    	 &3.07e+01  &2.71e+01 &2.58e+01 &2.62e+01\\
  convergence  	 &5.57e-04  &8.01e-04 &8.97e-04 &2.35e-03\\
  Total   	 &7.30e+02  &5.56e+02 &4.93e+02 &4.42e+02\\
  \hline
  \end{tabular}
\end{table}

\section{Discussions}

We highlight that different medium homogenization methods can result in different level of accuracy. The result obtained in this paper uses the homogenization method in \citep{davydycheva2003efficient}. The interface may also be handled using anti-aliasing bandlimited step function \citep{mittet2021small}, which can be very accurate also \citep{Yang_2023_HFDNU}. The step function approach is not adopted in \verb|libEMM|, because a continuous description of the 2D interface (usually difficult to have in practice) is required for modelling in 3D inhomogeneous medium.

Besides the accuracy, the efficiency is also another important concern. The author have reported a speedup factor of 40 \citep{Yang_2021_GPU_CSEM} using GPU compared to single threaded GPU. After applying many effective optimizations in the C code, in particular eliminating point-by-point convergence check on grid, the speedup factor becomes smaller, but still significant. Note that we rely on a GPU card purchased in 2014 which may be quite old-fashioned. A more impressive number on speedup should be expected using latest hardware. However, releasing the software is more worthwhile as it opens the door for the user to further improve the performance rather than focusing on specific number on a specific hardware.

There are some limitations in current version of \verb|libEMM|.
Modelling with extended source of finite length has also been implemented in \verb|libEMM| by uniformly distributing source current over the length of the the antenna. However, the accuracy of the implementation remains unknown. The FDTD approach is capable to handle fully anisotropic medium for numerical simulation. These obviously imply important applications for practical problems which have not been implemented in \verb|libEMM|. A simple tool to build 3D models of 1D and 2D structures has been served, but splitted from the modelling as an independent part. The users may use other advanced and fancy tools to build sophisticated 3D model with geological structures, and then perform medium homogenization over nonuniform grid based on the tools provided before starting modelling jobs.

\section{Conclusions}

A light weight yet powerful software - \verb|libEMM| - has been presented to do 3D CSEM modelling in fictitious wave domain. It includes both CPU implementation and GPU parallelization. Thus, \verb|libEMM| allows the users benefiting from both multi-core parallel cluster, but also promises efficiency improvement through graphics cards when GPU resources are available. A detailed description of the software features are given for the ease of usage. Application examples are given with performance analysis. The landscape of this work is to move this implementation on cluster for 3D CSEM inversion, to effectively handle large scale applications.

\section*{Acknowledgements}

Pengliang Yang is supported by Chinese Fundamental Research Funds for the Central
Universities (AUGA5710010121) and National Natural Science Fundation of China (42274156). The author thanks Rune Mittet and Daniel Shantsev for many fruitful discussions. The author also appreciates the thoughtful question on the analysis of the dispersion error raised by an anonymous reviewer. \verb|libEMM| is available via github repository: \url{https://github.com/yangpl/libEMM}. Many user-friendly features and development ideas of \verb|libEMM| are inspired by the well-known computational geophysics open softwares - \verb|Seismic Unix| and \verb|Madagascar|.



\appendix
\section{Dispersion analysis}

Let us present some theoretic analysis on the numeric dispersion in wave domain without source terms. Applying the curl operator for the first subequation, and then plugging it into the second one in equation \eqref{eq:mittet} cancels out the electric field such that
\begin{equation}\label{eq:disp}
(\mu\epsilon \partial_{tt} + \nabla\times\nabla\times)\mathbf{H}'
=(\mu\epsilon \partial_{tt} - \Delta)\mathbf{H}'=0,
\end{equation}
where $\Delta=\nabla\cdot \nabla$; the second equality is due to $\nabla \cdot\mathbf{H} = 0$ and the fact that
\begin{equation}\label{eq:curlcurl}
  \nabla\times\nabla\times\mathbf{H}'=\nabla\nabla\cdot\mathbf{H}'-\nabla\cdot\nabla\mathbf{H}'.
\end{equation}
Consider the time harmonic plane wave representation of the EM fields on the grid 
\begin{equation}
  \mathbf{H}' \propto e^{-\mathrm{i}(\omega t - k_x  x - k_y  y - k_z z)},
\end{equation}
where $k_x$, $k_y$ and $k_z$ are the wavenumber in the x, y and z directions.  In case of uniform grid spacing $\Delta x$, $\Delta y$ and $\Delta z$, the grids are then directly prescribed by the index: $x_i=i\Delta x$, $y_i=i\Delta y$ and $z_i=i\Delta z$. The $2r$ coefficients of the finite difference stencil are independent of the grid position, satisfying 
\begin{equation}
  c_i^+ = -c_{-i+1}^+ =  c_i^- =-c_{-i+1}^-:=\frac{c_i}{\Delta x}, \quad i=1,\cdots, r
\end{equation}
where $c_i$ is the standard finite difference coefficients which can be easily found using the method in \citep{Fornberg_1998_NFD}. The discretized spatial derivatives become
\begin{equation}
  D_x^+=D_x^-=\frac{1}{\Delta x}\sum_{i=1}^r c_i (e^{\mathrm{i}\frac{(2i-1)k_x\Delta x}{2}}-e^{-\mathrm{i}\frac{(2i-1)k_x\Delta x}{2}})
  =\frac{2\mathrm{i}}{\Delta x}\sum_{i=1}^r c_i \sin((i-\frac{1}{2})k_x\Delta x),
\end{equation}
leading to the discretized Laplacian operator
\begin{equation}
  \begin{split}
  \Delta \approx &D_x^+D_x^- + D_y^+D_y^- + D_z^+D_z^-\\
  =& -\frac{(2\sum_{i=1}^r c_i \sin((i-\frac{1}{2})k_x\Delta x))^2}{\Delta x^2}
  -\frac{(2\sum_{i=1}^r c_i \sin((i-\frac{1}{2})k_y\Delta y))^2}{\Delta y^2}
  -\frac{(2\sum_{i=1}^r c_i \sin((i-\frac{1}{2})k_z\Delta z))^2}{\Delta z^2}\\
  \end{split}
\end{equation}
The time derivative was approximated using 2nd order leap-frog scheme so that
\begin{equation}
\partial_{tt}\approx D_t^+D_t^-=-\left(\frac{2\sin(\frac{\omega\Delta t}{2})}{\Delta t} \right)^2.
\end{equation}
Equation \eqref{eq:disp} results in
\begin{equation}\label{eq:disp2}
  \frac{\sin^2(\frac{\omega\Delta t}{2})}{v^2\Delta t^2} 
  =\frac{(\sum_{i=1}^r c_i \sin((i-\frac{1}{2})k_x\Delta x))^2}{\Delta x^2}
  +\frac{(\sum_{i=1}^r c_i \sin((i-\frac{1}{2})k_y\Delta y))^2}{\Delta y^2}
  +\frac{(\sum_{i=1}^r c_i \sin((i-\frac{1}{2})k_z\Delta z))^2}{\Delta z^2}.
\end{equation}

Denote
\begin{equation}
  \begin{array}{rl}
S:=&v\Delta t\sqrt{\frac{(\sum_{i=1}^r c_i \sin((i-\frac{1}{2})k_x\Delta x))^2}{\Delta x^2}
  +\frac{(\sum_{i=1}^r c_i \sin((i-\frac{1}{2})k_y\Delta y))^2}{\Delta y^2}
  +\frac{(\sum_{i=1}^r c_i \sin((i-\frac{1}{2})k_z\Delta z))^2}{\Delta z^2}}
\end{array}
\end{equation}
and the magnitude of the wavenumber $k:=\sqrt{k_x^2+k_y^2+k_z^2}$. According to $k=\omega/v$, equation \eqref{eq:disp2} implies the numerical phase velocity $v_{phase}$ in FDTD simulation satisfies
\begin{equation}
  \sin(\frac{k\Delta t v_{phase}}{2})= S.
\end{equation}
The dispersion error can then be defined by the deviation of the relative phasevelocity (the ratio between numeric phase velocity and the true velocity) from 1,
\begin{equation}
\delta := \frac{v_{phase}}{v}-1=\frac{2}{kv\Delta t}\arcsin S-1.
\end{equation}
The 1D analysis for wave equation on uniform grid can be found in \citet[chapter 2]{Taflove_2005_CEF}. Indeed, \citet{Dablain_1986_AHO} shows that the use of higher order finite difference scheme reduces the dispersion error, allowing simulation with less number of points per wavelength. For 3D Maxwell equation, the mathematical derivation above gives a qualitative way for such analysis. Unfortunately, this analysis becomes challenging on nonuniform grid.

\bibliographystyle{apalike} 
\newcommand{\SortNoop}[1]{}

\end{document}